\documentclass[copyright,creativecommons]{eptcs}
\providecommand{\event}{QAPL 2012} % Name of the event you are submitting to

\usepackage{breakurl}             % Not needed if you use pdflatex only.
\usepackage{amsmath}
\usepackage{amssymb}
\usepackage[utf8]{inputenc}
\usepackage{xspace,graphicx}
\usepackage{textcomp} %used for special symbols
\usepackage{calc}
\usepackage{multicol}
\usepackage[draft]{fixme} %set to final before submission
\usepackage{wrapfig}
\usepackage[british]{babel}
\usepackage{booktabs}

\newcommand{\WMITL}{\ensuremath{\text{WMTL}_\leq}}
\newcommand{\uppaal}{{\sc Uppaal}\xspace}
\newcommand{\uppaalsmc}{{\sc Uppaal-smc}\xspace}
\newcommand{\tiga}{{\sc Uppaal-tiga}\xspace}
\newcommand{\U}{\ensuremath{\text{\sf{U}}}}
\newcommand{\OO}{\ensuremath{\text{\sf{O}}}}

\newcommand{\arrow}[1]{\stackrel{#1}{\longrightarrow}}

\title{\uppaalsmc: \\Statistical Model Checking for Priced Timed Automata
\thanks{
The paper is supported by VKR Centre of Excellence -- MT-LAB and the
IDEA4CPS center established on a grant from Danish National Research
Foundation
}
}
\author{
Peter~Bulychev\quad
Alexandre~David\quad
Kim~Guldstrand~Larsen\\
Marius~Mikučionis\quad
Danny~Bøgsted~Poulsen
\institute{Department of Computer Science\\
Aalborg University, Denmark}
\email{\{pbulychev,adavid,kgl,marius,dannybp\}@cs.aau.dk}
\and
Axel~Legay
\institute{INRIA Rennes, France\\
Department of Computer Science\\
Aalborg University, Denmark}
\email{alegay@irisa.fr}
\and
Zheng~Wang
\institute{Shanghai Key Laboratory of Trustworthy Computing\\
Software Engineering Institute\\
 East China Normal University, China}
}

\def\titlerunning{Statistical  Model  Checking for  Priced
  Timed     Automata}     \def\authorrunning{A. Legay et al.}

\begin{document}
\maketitle

\begin{abstract}
  This paper offers  a survey of \uppaalsmc, a major extension of
  the real-time  verification tool \uppaal. \uppaalsmc  allows for the
  efficient analysis  of performance  properties of networks  of priced
  timed automata under a natural stochastic semantics.  In particular,
  \uppaalsmc relies on a series of extensions of the statistical model
  checking  approach  generalized  to  handle  real-time  systems  and
  estimate  undecidable problems.   \uppaalsmc comes  together  with a
  friendly  user  interface that  allows  a  user  to specify  complex
  problems in  an efficient manner as  well as to get  feedback in the
  form  of  probability  distributions  and compare  probabilities  to
  analyze performance aspects of systems.   The focus of the survey is
  on the evolution of the tool -- including modeling and specification
  formalisms   as  well   as  techniques   applied  --   together  with
  applications of the tool to case studies.
\end{abstract}

\section{Introduction}

Quantitative properties of stochastic systems are usually specified in
logics that allow one to compare the measure of executions satisfying
certain temporal properties with thresholds. The model checking
problem for stochastic systems with respect to such logics is
typically solved by a numerical approach~\cite{BHHK03,CG04} that
iteratively computes (or approximates) the exact measure of paths
satisfying relevant sub-formulas; the algorithms themselves depend on
the class of systems being analyzed as well as the logic used for
specifying the properties.

Another approach to solve the model checking problem is to
\emph{simulate} the system for finitely many runs, and use
\emph{hypothesis testing} to infer whether the samples provide a
\emph{statistical} evidence for the satisfaction or violation of the
specification~\cite{You05a}. The crux of this approach is that since
sample runs of a stochastic system are drawn according to the
distribution defined by the system, they can be used to get estimates
of the probability measure on executions.  Those techniques, also
called {\em Statistical Model Checking techniques}
(SMC)\,\cite{HLMP04,SVA04,You05a,ODGLP11}, can be seen as a trade-off
between testing and formal verification. In fact, SMC is very similar
to Monte Carlo used in industry, but it relies on a formal model of
the system. The core idea of SMC is to monitor a number of simulations
of a system whose behaviors depend on a stochastic semantic. Then, one
uses the results of statistics (e.g. sequential hypothesis testing or
Monte Carlo) together with the simulations to get an overall estimate
of the probability that the system will behave in some manner. While
the idea resembles the one of classical Monte Carlo simulation, it is
based on a formal semantic of systems that allows us to reason on very
complex behavioral properties of systems (hence the terminology). This
includes classical reachability properties such as ``can I reach such a
state?'', but also non trivial properties such as ``can I reach this
state x times in less than y units of time?''. Of course, in contrast
with an exhaustive approach, such a simulation-based solution does not
guarantee a result with 100\% confidence. However, it is possible to
bound the probability of making an error. Simulation-based methods are
known to be far less memory and time intensive than exhaustive ones,
and are sometimes the only option\,\cite{YKNP06,JKOSZ07}.

Statistical model checking is now widely accepted in various research
areas such as software engineering, in particular for industrial
applications\,\cite{BBBCDL10,MPL11,CZ11}, or even for solving problems
originating from systems biology\,\cite{CFLHJL08,JCLLPZ09}. There are
several reasons for this success.  First, SMC is very simple to
understand, implement, and use.  Second, it does not require extra
modeling or specification effort, but simply an operational model of
the system, that can be simulated and checked against state-based
properties.  Third, it allows us to verify
properties\,\cite{CDL08,CDL09,BBBCDL10} that cannot be expressed in
classical temporal logics. Finally, SMC allows to approximate
undecidable problems. This latter observation is crucial. Indeed most
of emerging problems such as energy consumption are
undecidable\,\cite{FJLS11,BFLM10} and can hence only be estimated. SMC
has been applied to a wide range of problems that goes from embedded
systems\cite{CDL08} and systems biology\,\cite{CDL08,CDL09} to more
industrial applications\,\cite{BBBCDL10}.

In a series of recent works\,\cite{dllmw11,bdlml11,dllmpvw11}, we have
investigated the problem of Statistical Model Checking for networks
of Priced Timed Automata (PTA). PTAs are timed automata, whose clocks
can evolve with different rates, while\footnote{in contrast to the
  usual restriction of priced timed automata \cite{PTA01,WTA01}} being
used with no restrictions in guards and invariants. In \cite{dllmpvw11}, we
have proposed a natural stochastic semantic for such automata, which
allows to perform statistical model checking. Our work has later been
implemented in \uppaalsmc, that is a stochastic and statistical model
checking extension of \uppaal. \uppaalsmc relies on a series of
extensions of the statistical model checking approach generalized to
handle real-time systems and estimate undecidable problems.
\uppaalsmc comes together with a friendly user interface that allows a
user to specify complex problems in an efficient manner as well as to
get feedback in the form of probability distributions and compare
probabilities to analyze performance aspects of systems.

The objective of this paper is to offer a survey of \uppaalsmc. This
includes modeling and specification formalism as well as techniques
applied -- together with applications of the tool to case studies.

\paragraph{\bf Structure of the paper}

In Section \ref{sec:formalism}, we introduce the formalism of networks
of Priced timed automata. Section \ref{sec:queries} provides an overview of some
existing statistical model checking algorithms, while Sections
\ref{sec:gui} and \ref{sec:engine} introduce the GUI and give some
details on the engine of \uppaalsmc.
Finally, Section \ref{sec:cases} presents a series of applications for
the tool-set and Section \ref{sec:conclusion} concludes the paper.

\section{Modeling Formalism}
\label{sec:formalism}

The new  engine of \uppaalsmc \cite{dllmw11} supports  the analysis of
Priced Timed Automata (PTAs) that  are timed automata whose clocks can
evolve  with different  rates  in different  locations.  In fact,  the
expressive power  (up to  timed bisimilarity) of  NPTA equals  that of
general  linear hybrid  automata (LHA)  \cite{ACHH95},  rendering most
problems -- including that of reachability -- undecidable.
%\uppaal models can also contain integer variables that can be present in transition guards, and they
%can be updated only when a discrete transition is taken.
We  also   assume  PTAs  are  input-enabled,   deterministic  (with  a
probability measure defined on  the sets of successors), and non-zeno.
PTAs  communicate  via  broadcast  channels and  shared  variables  to
generate Networks of Price Timed Automata (NPTA).

 \begin{wrapfigure}{r}{0.3\linewidth}
   \centering
%   \vspace{-25pt}
   \begin{tabular}{@{}*{3}{c}@{}}
     \includegraphics[height=0.15\textheight]{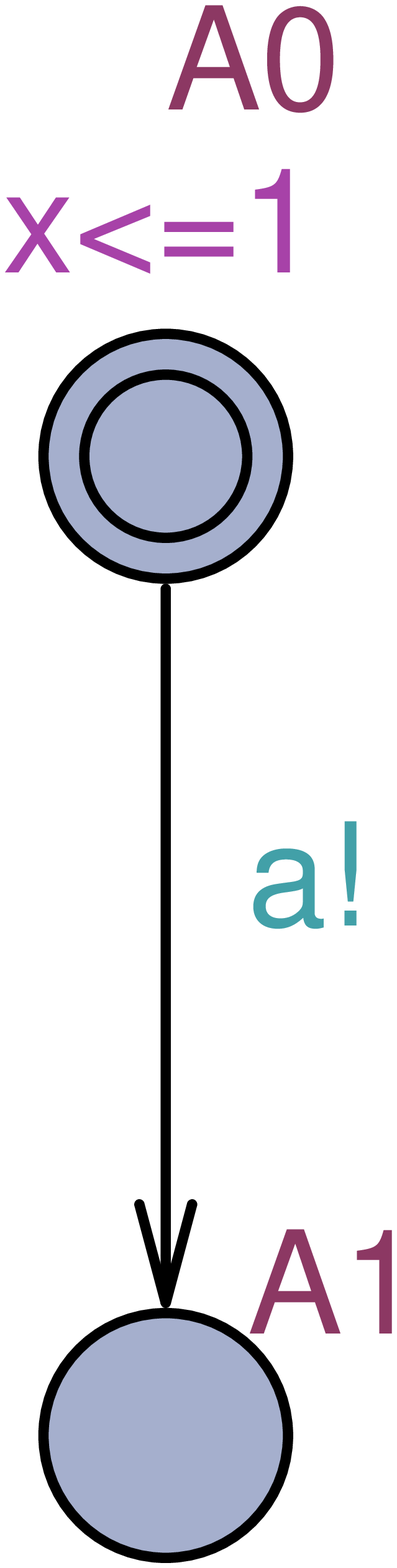} &
     \includegraphics[height=0.15\textheight]{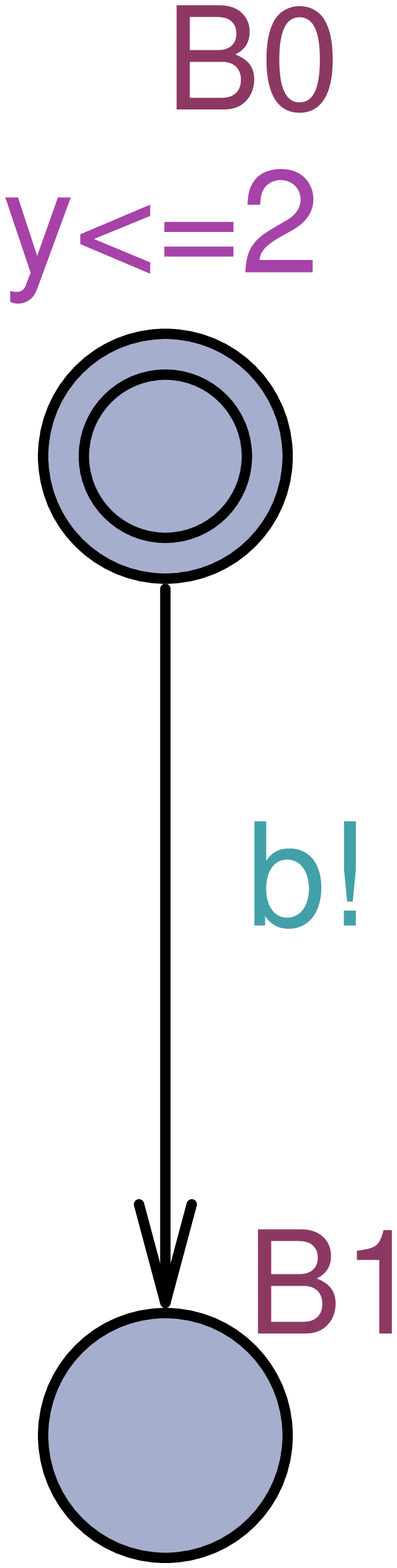} &
     \includegraphics[height=0.15\textheight]{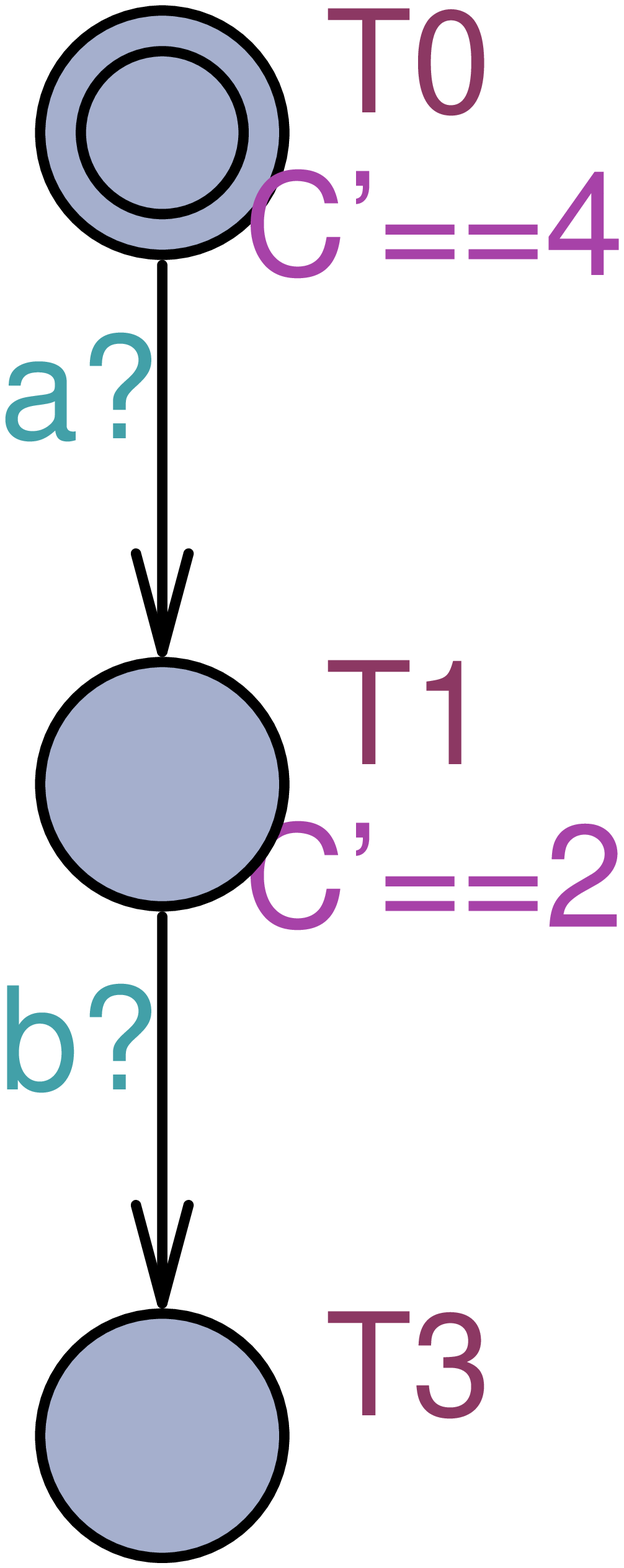} \\
     $A$ & $B$ &  $T$ \\
   \end{tabular}
   \vspace{-5pt}
   \caption{An NPTA, $( A|B|T)$.}
  \label{fig:NPTA}
  \vspace{-15pt}
 \end{wrapfigure}

%  \begin{figure}[t]
%   \centering
%  \begin{tabular}{*{3}{c}}
%      \includegraphics[height=0.15\textheight]{A} &
%     \includegraphics[height=0.15\textheight]{B} &
%      \includegraphics[height=0.15\textheight]{T} \\
%      $A$ & $B$ &  $T$ 
% \end{tabular}\,
%     \includegraphics[width=0.4\textwidth]{simpleplot.eps}
%     \caption{An  NPTA, $(  A|B|T)$ with  cumulative  probabilities for
%       {\tt  time} and  {\tt cost}-bounded  reachability  of $T_3$.}
%    \label{fig:NPTA}
% \end{figure}

Fig.~\ref{fig:NPTA} provides  an NPTA with three  components $A$, $B$,
and $T$  as specified using the  \uppaal GUI. One can  easily see that
the composite system $(A|B|T)$ has  the transition sequence:
 \smallskip \\
{\small $
  \big((A_0,B_o,T_0),[x=0,y=0,C=0]\big) \arrow{1}\arrow{a!}\\
%\big((A_0,B_0,T_0),[x=1,y=1,C=4]\big) \arrow{a!}\\
\big((A_1,B_0,T_1),[x=1,y=1,C=4]\big) \arrow{1} \arrow{b!}\\
%\big((A_1,B_0,T_1),[x=2,y=2,C=6]\big) \arrow{b!}
\big((A_1,B_1,T_2),[x=2,y=2,C=6]\big),$}\smallskip\\
%\end{eqnarray*}
demonstrating that the final location $T_3$ of $T$ is reachable.
In fact, location $T_3$ is reachable within cost $0$ to $6$ and within
total time  $0$ and $2$ in  $(A|B|T)$ depending on when  (and in which
order)  $A$ and $B$  choose to  perform the  output actions  $a!$ and
$b!$.  Assuming  that the choice  of these time-delays is  governed by
probability  distributions, a  measure on  sets  of runs  of NPTAs  is
induced,   according  to   which  quantitative   properties   such  as
\emph{``the  probability  of  $T_3$   being  reached  within  a  total
  cost-bound of 4.3''} become well-defined.

\begin{wrapfigure}{l}{0.45\linewidth}
  \vspace*{-5pt}
  \includegraphics[width=\linewidth]{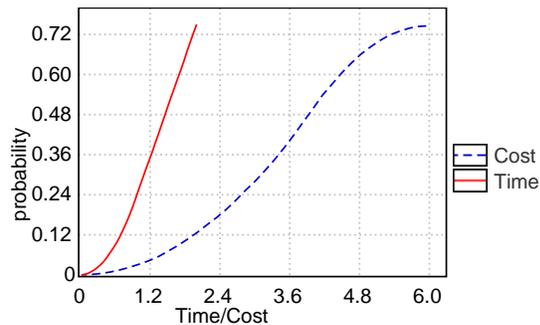}
  \vspace*{-15pt}
  \caption{Cumulative   probabilities   for    {\tt   time}   and   {\tt
    Cost}-bounded reachability of $T_3$.}
  \label{CumTC:fig}
\end{wrapfigure}
In  our  early  works  \cite{dllmpvw11},  we  provide  a  natural
stochastic   semantics, where PTA   components  associate   probability
distributions to  both the time-delays  spent  in a given  state as
well  as to the  transition between  states.  In \uppaalsmc  uniform
distributions   are  applied  for   bounded  delays   and  exponential
distributions for  the case where a component  can remain indefinitely
in  a state.   In a  network of  PTAs the  components  repeatedly race
against each other, i.e.  they independently and stochastically decide
on their own how much  to delay before outputting, with the ``winner''
being the component that chooses  the minimum delay.  For instance, in
the NPTA  of Fig.~\ref{fig:NPTA}, $A$  wins the initial race  over $B$
with probability $0.75$.

As observed  in \cite{dllmpvw11}, though  the stochastic semantic
of  each individual  PTA  in \uppaalsmc is  rather  simple (but  quite
realistic), arbitrarily complex stochastic behavior can be obtained by
their composition when mixing individual distributions through message
passing.   The beauty  of our  model is  that these  distributions are
naturally and automatically defined by the network of PTAs.

\begin{figure*}[htb]
  \centering
  \begin{tabular}{ll}
    a) Axel &\includegraphics[height=0.11\textheight]{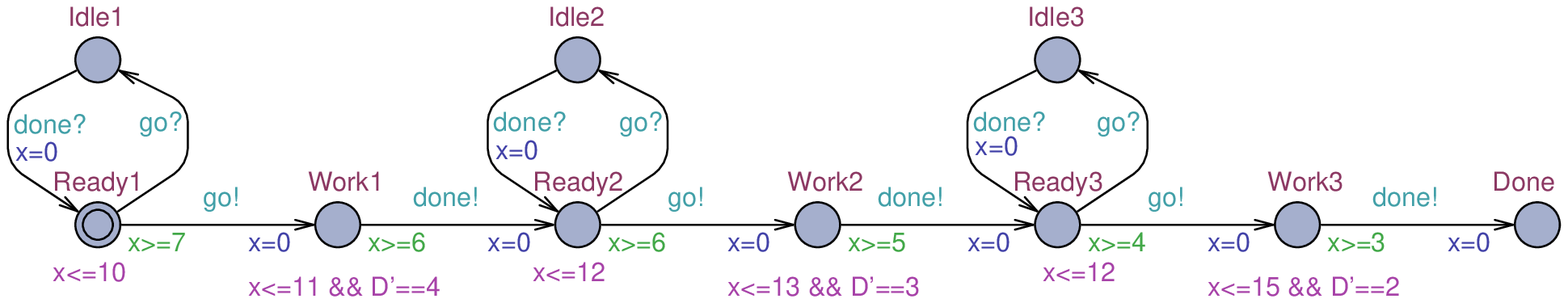}\\
    b) Alex &\includegraphics[height=0.11\textheight]{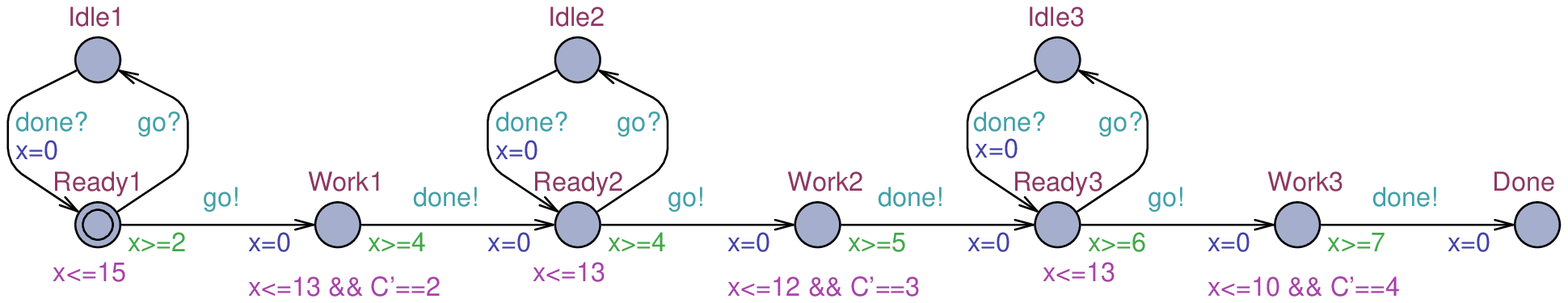} 
  \end{tabular}
  \vspace*{-10pt}
  \caption{3-Nail Hammer Game between Axel and  Alex.}
  \label{fig:hammer}
\end{figure*}

\paragraph{The Hammer Game}
To illustrate the stochastic semantics further consider the network of
two  priced   timed  automata  in   Fig.~\ref{fig:hammer}  modeling  a
competition  between the  two players  Axel  and Alex  both having  to
hammer three nails down.  As can be seen by
the representing {\tt Work}-locations the time (-interval) and rate of
energy-consumption required for hammering a nail depends on the player
and the  nail-number.  As  expected Axel is  initially quite  fast and
uses a lot of energy but  becomes slow towards the last nail, somewhat
in contrast to  Alex. To make it an  interesting competition, there is
only  \emph{one} hammer illustrated  by repeated  competitions between
the two players in the {\tt Ready}-locations, where the slowest player
has to  wait in  the {\tt Idle}-location  until the faster  player has
finished hammering the next nail.  Interestingly, despite the somewhat
different strategy applied, the  best- and worst-case completion times
are identical for Axel and Alex: 59 seconds and 150 seconds. So, there
is no  difference between  the two players  and their strategy,  or is
there?

Assume now that a third person wants  to bet on who is the more likely
winner  --  Axel or  Alex  -- given  a  refined  semantics, where  the
time-delay before performing an  output is chosen stochastically (e.g.
by  drawing from  a uniform  distribution) and  independently  by each
player  (component).   

\begin{figure*}[ht] 
  \begin{tabular}{@{}cc@{}}
    \includegraphics[width=0.48\textwidth]{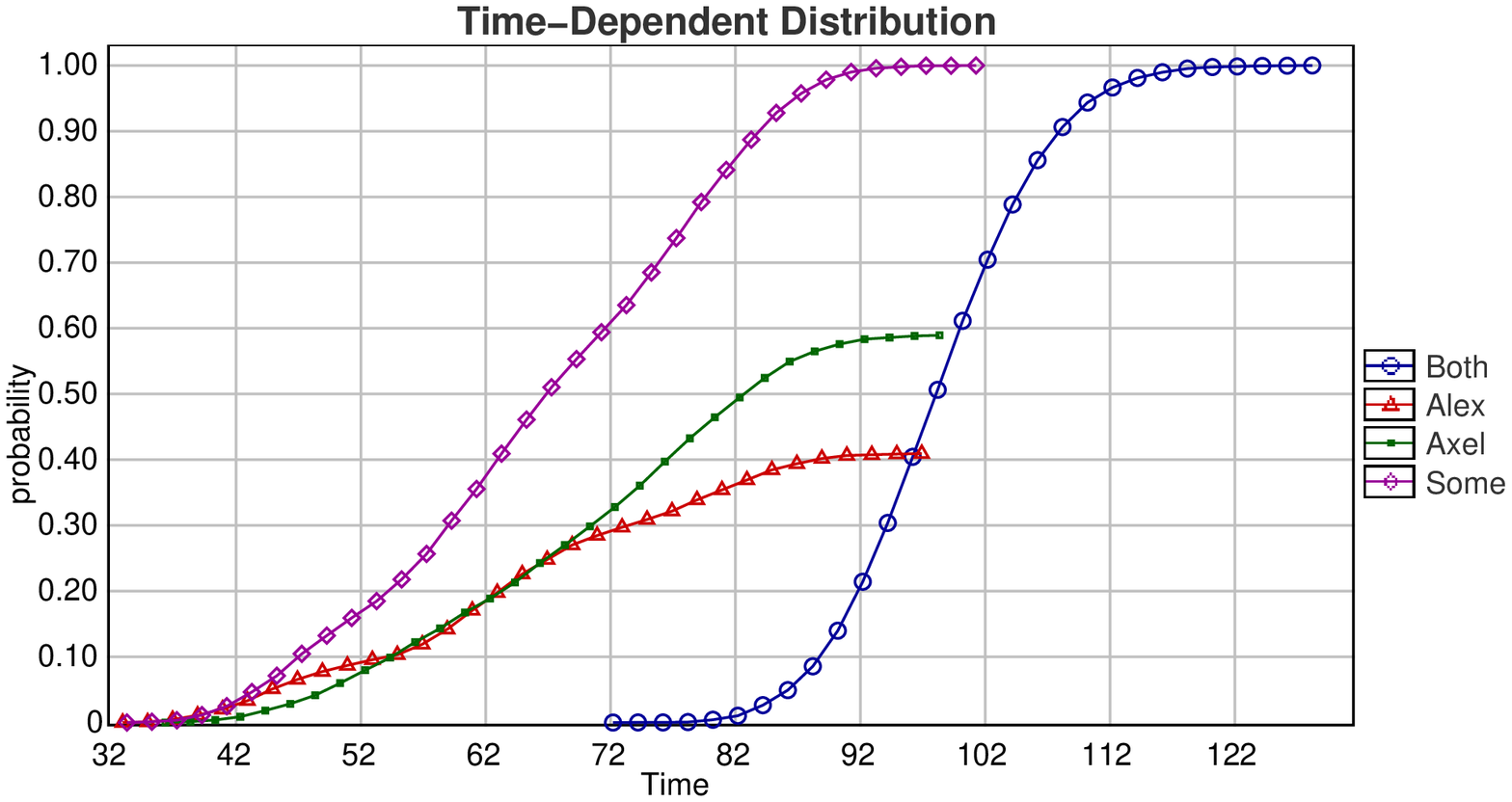} &
    \includegraphics[width=0.48\textwidth]{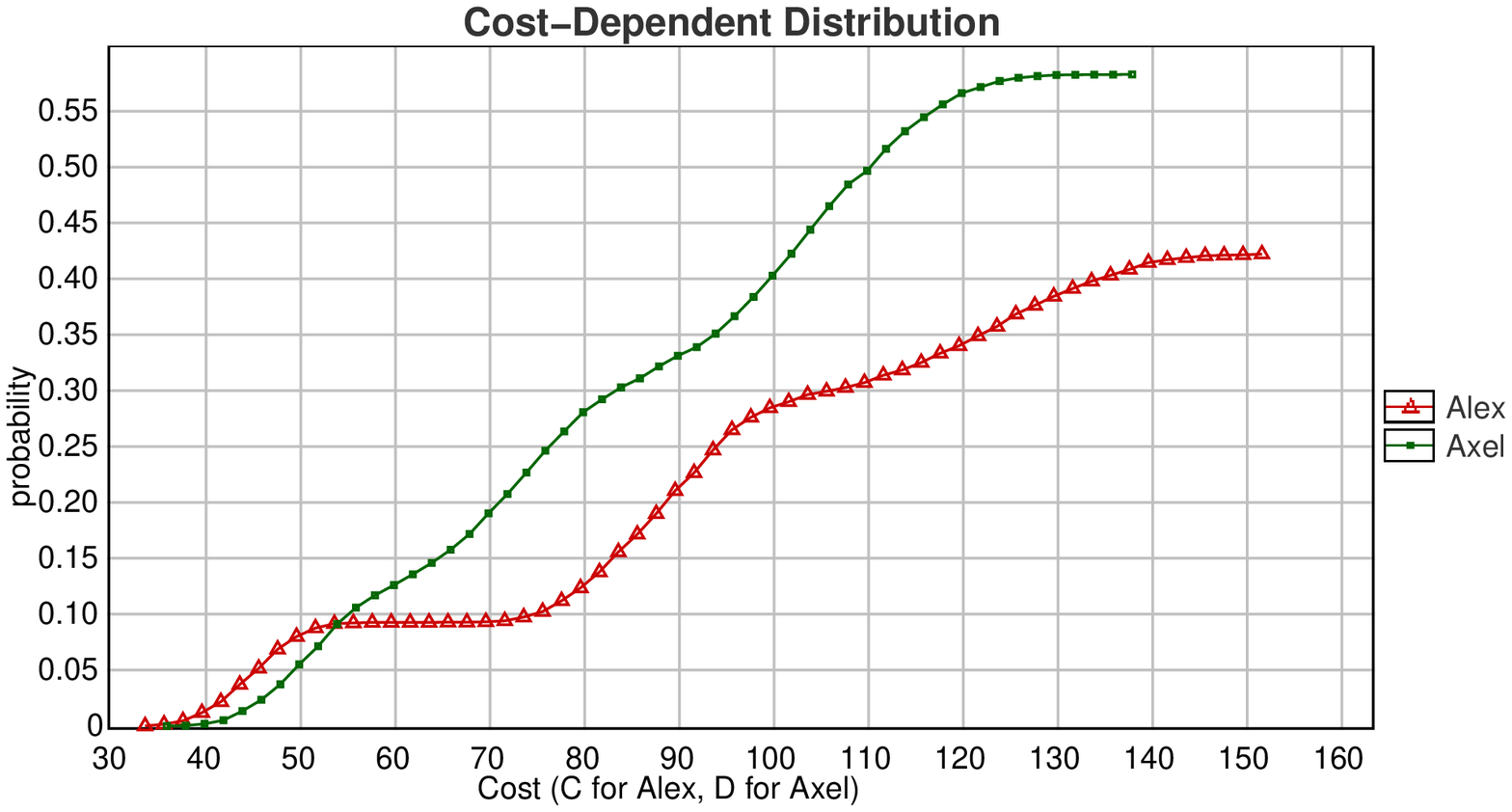} \\
    a) & b) \\
  \end{tabular}
  \vspace*{-8pt}
  \caption{Time- and Cost-dependent Probability of winning the
    Hammer Game}
\label{fig:TimeCost}
\end{figure*}

Under  such a  refined  semantics there  is  a significant  difference
between  the two  players  (Axel and  Alex)  in the  Hammer Game.   In
Fig.~\ref{fig:TimeCost}a) the probability  distributions for either of
the two players winning before a  certain time is given.  Though it is
clear that  Axel has a higher  probability of winning  than Alex (59\%
versus 41\%) given unbounded time, declaring the competition  a draw if it
has not finished before 50 seconds actually makes Alex the more likely
winner.    Similarly,    Fig.~\ref{fig:TimeCost}b)   illustrates   the
probability of either of the  two players winning given an upper bound
on energy.   With an unlimited amount  of energy, clearly  Axel is the
most  likely winner,  whereas limiting  the consumption  of  energy to
maximum 52 ``energy-units'' gives Alex an advantage.

\paragraph{Extended Input Language}
\uppaalsmc  takes as  input NPTAs  as described  above.  Additionally,
there is  support for  other features of  the \uppaal  model checker's
input  language  such  as   integer  variables,  data  structures  and
user-defined  functions,  which  greatly ease  modeling.   \uppaalsmc
allows the user to specify  an arbitrary (integer) rate for the clocks
on  any location. In  addition, the  automata support  branching edges
where  weights  can  be  added  to give  a  distribution  on  discrete
transitions. It  is important  to note that  rates and weights  may be
general  expressions that  depend on  the states  and not  just simple
constants.

\begin{figure}[htb]
\centering
\begin{tabular}{c@{\hspace{1cm}}c}
\includegraphics[width=0.35\linewidth]{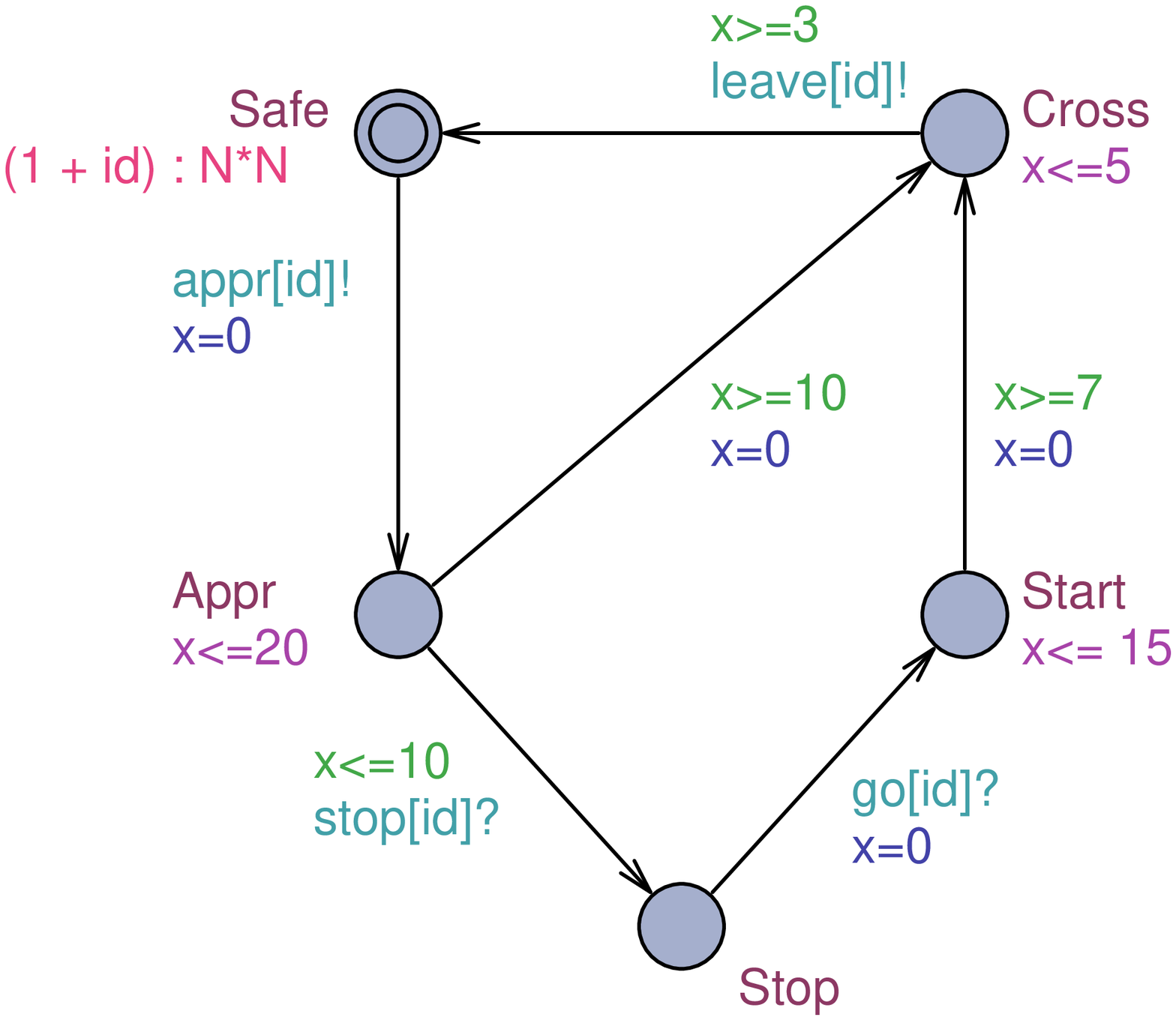}&
\includegraphics[width=0.31\linewidth]{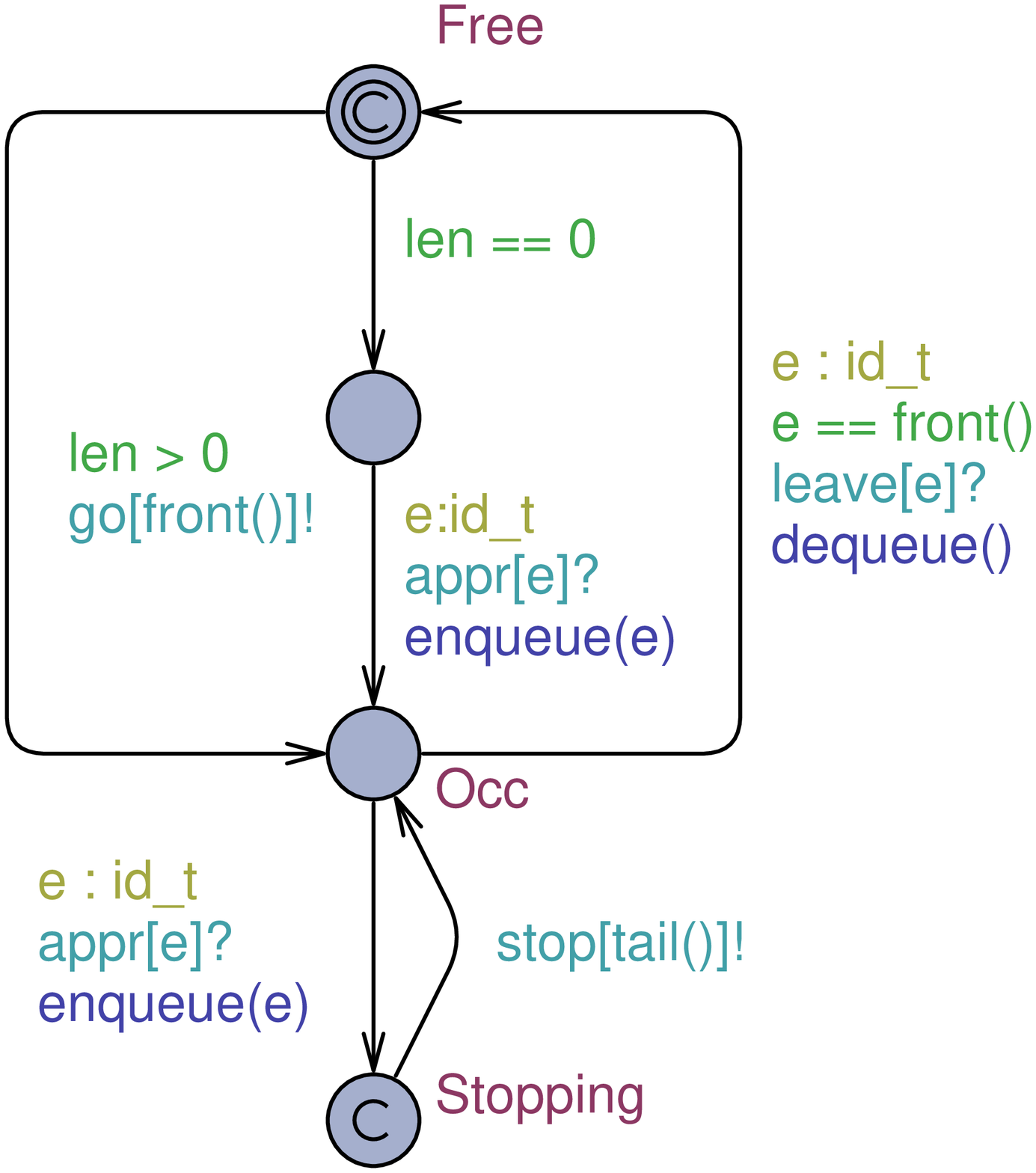} \\
(a) & (b)
\end{tabular}
\caption{Template of a Train (a) and the Gate Controller (b).}
\label{fig:train}
\end{figure}

To illustrate the extended input language, we consider a train-gate
example. This example is available in the distributed version of
\uppaalsmc.
A number of trains are approaching a bridge on which there is only one
track. To avoid collisions, a controller stops the trains. It restarts
them when possible to make sure that trains will eventually cross the
bridge. There are timing constraints for stopping the trains modeling
the fact that it is not possible to stop trains instantly. The
interesting point w.r.t. SMC is to define the arrival rates of these
trains. Figure~\ref{fig:train}(a) shows the template for a train. The
location \texttt{Safe} has no invariant and defines the rate of the
exponential distribution for delays. Trains delay according to this
distribution and then approach and synchronize with \texttt{appr[i]!}
with the gate controller. Here we
define the rational $\frac{1+id}{N^2}$ where $id$ is the identifier of
the train and $N$ the number of trains. Rates are given by expressions
that can depend on the current states. Trains with higher $id$ arrive
faster. Taking transitions from locations with invariants is given by a
uniform distribution. This happens in \texttt{Appr}, \texttt{Cross}, and
\texttt{Start}, e.g., it takes some time picked uniformly between 3 and
5 time units to cross the bridge. Figure~\ref{fig:train}(b) shows the
gate controller that keeps track of the trains with an internal queue
data-structure (not shown here). It uses functions to queue trains
(when a train is approaching while the bridge is occupied in
\texttt{Occ}) or dequeue them when possible (when the bridge is free and
some train is queued).

% The  foundation of  \uppaalsmc was  presented in  \cite{dllmpvw11}, in
% particular  its stochastic  semantics and  different types  of checks,
% namely  probability evaluation,  hypothesis  testing, and  probability
% comparison.  In \cite{dllmw11}  the  tool was  presented. Both  papers
% showed the usability and  performance of the tool, comparing favorably
% against Prism and Pvesta.

\begin{figure}[!htb]
  \begin{tabular}{@{}c@{}c@{}c@{}}
    \includegraphics[height=0.125\textheight]{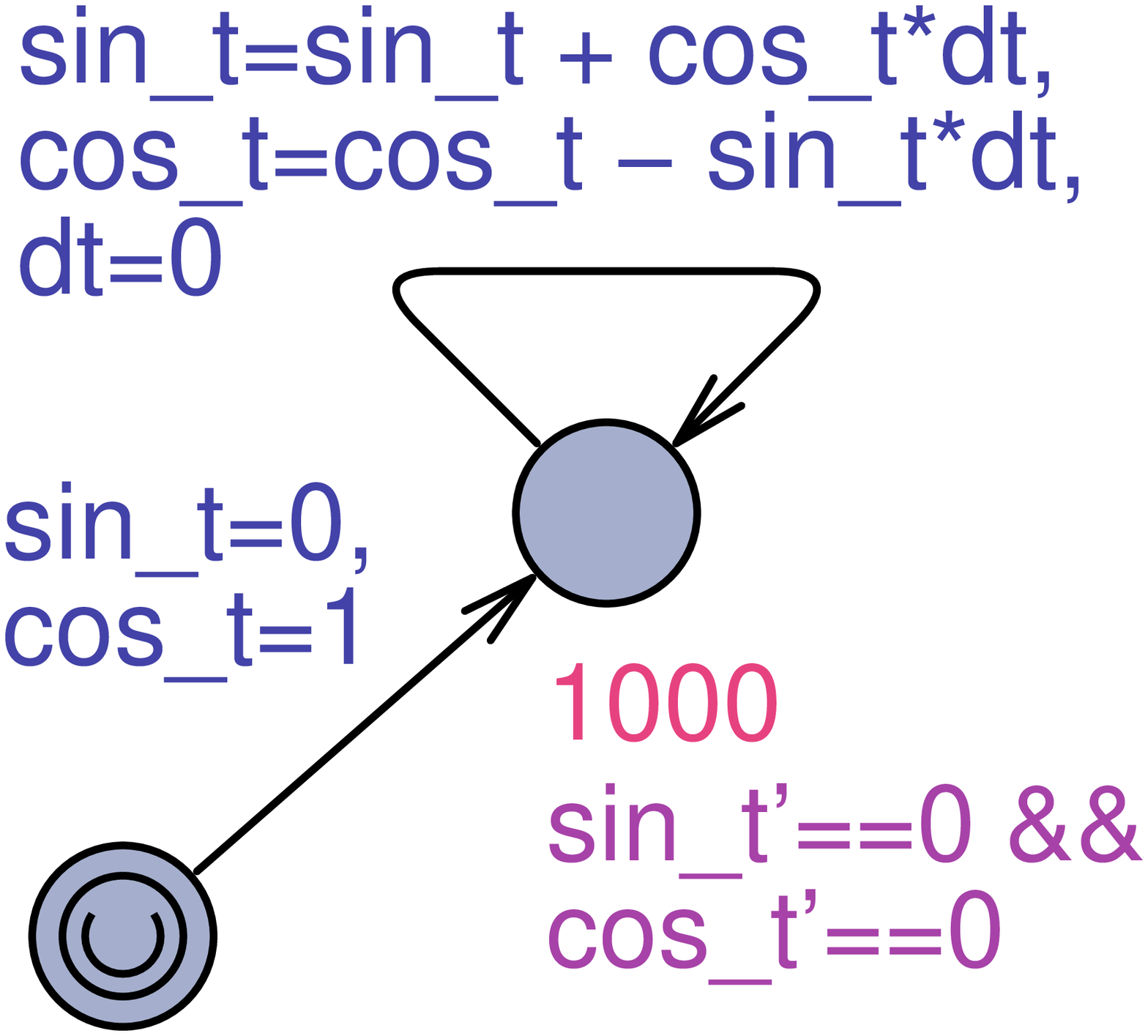} &
    \includegraphics[height=0.125\textheight]{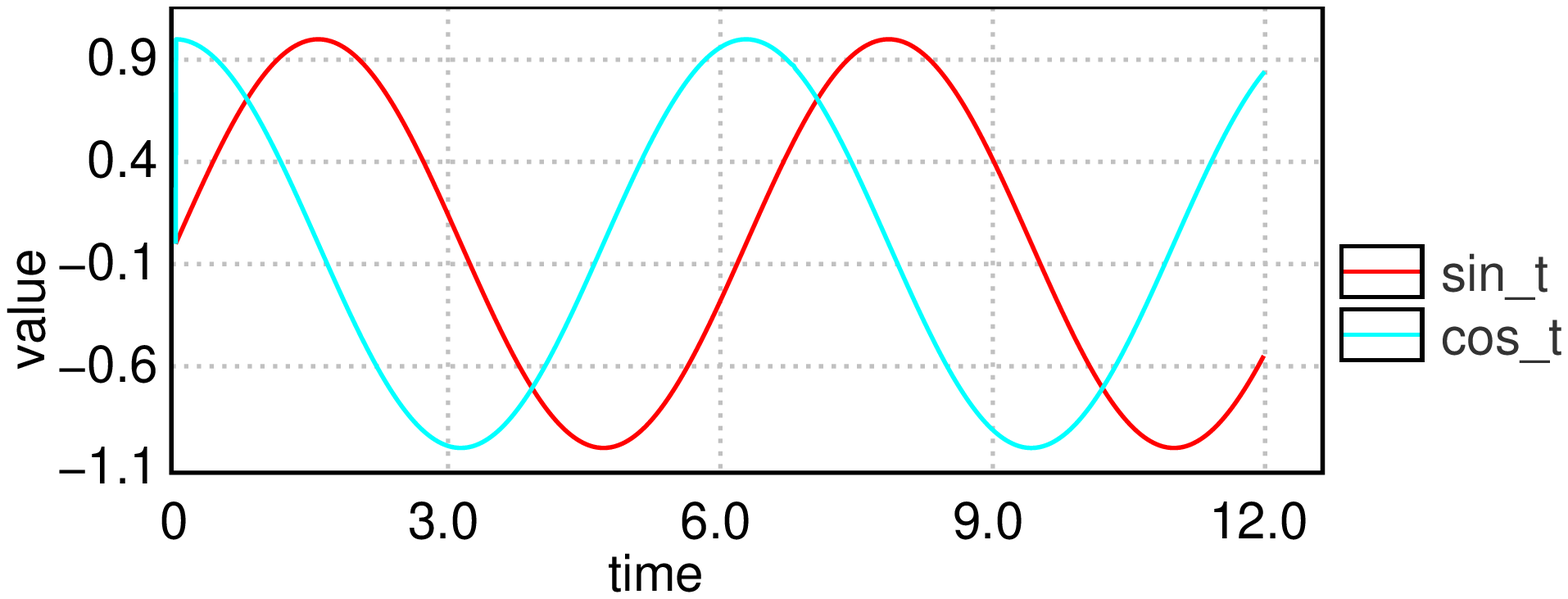} &
    \includegraphics[height=0.125\textheight]{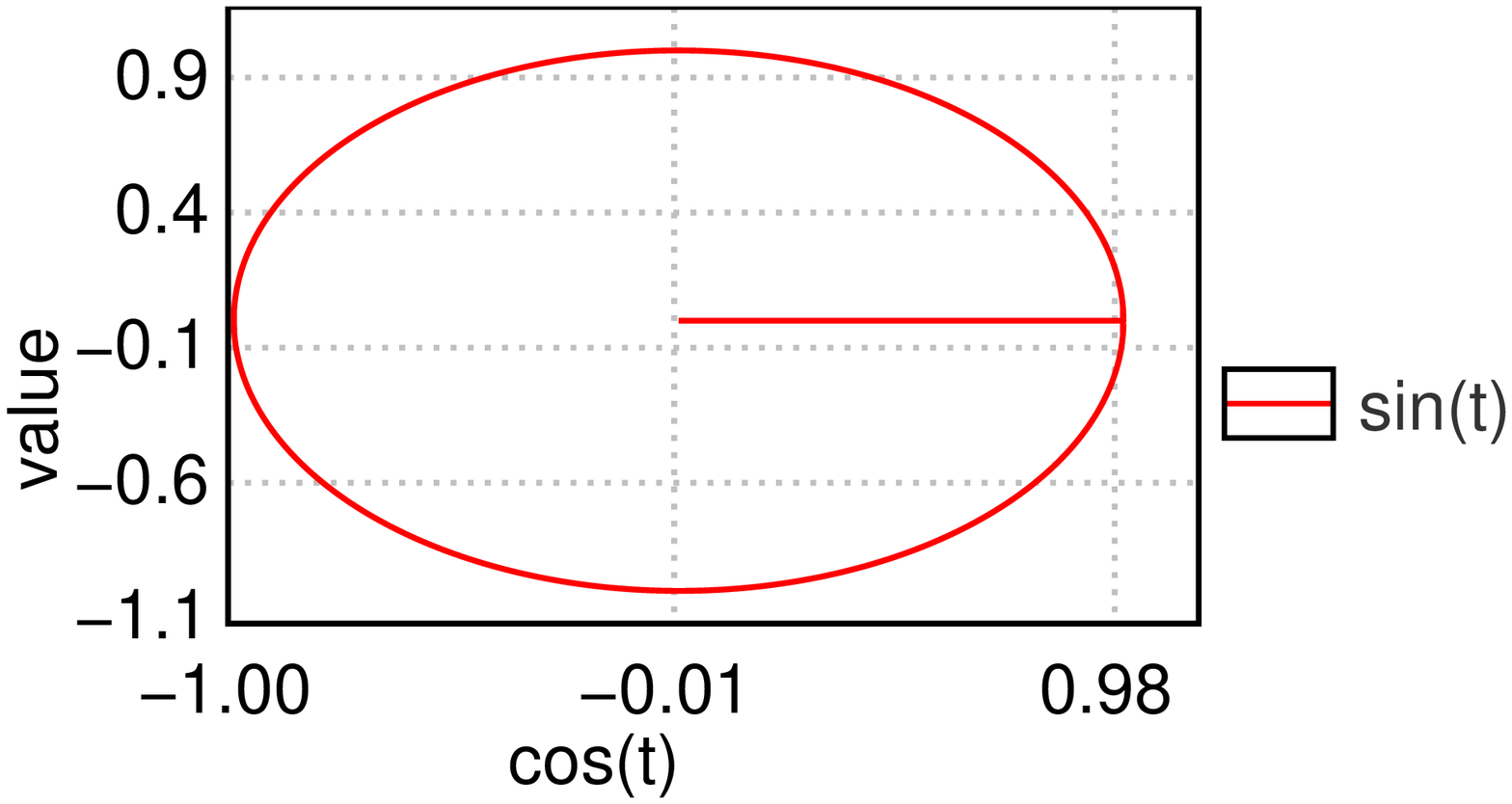}\\
    (a) & (b) & (c)
  \end{tabular}
  \vspace*{-8pt}
  \caption{How to use clock arithmetic to integrate complex functions.}
  \label{fig:cos}
\end{figure}

\paragraph{Floating Point Arithmetic}
For  modeling  certain  systems,  e.g., biological  systems,  integer
arithmetic  shows  its precision  limits  very  quickly.  The  current
engine implements  simple arithmetic operations on  clocks as floating
point variables.   This allows various tricks, in  particular the tool
can compute  nontrivial functions  using small step  integration.  For
example, Figure~\ref{fig:cos}(a) shows a timed automaton with floating
point arithmetic.
The  clocks \texttt{sin\_t}  and \texttt{cos\_t}  are used  to compute
$sin(t)$  and   $cos(t)$  using  simple   facts  as  $sin(t+dt)\approx
sin(t)+sin'(t)dt$  for  small  steps  of  $dt\rightarrow  0$,  whereas
$sin'(t)=cos(t)$  and  $sin(0)=0$, and  similarly  for $cos(t)$.   The
interesting trick  on the  model is the  high exponential  rate (1000)
that tells the engine to take small (random) time steps and record the
duration in clock {\tt dt}.   The other clocks are stopped and updated
on transition.  The value evolution of variables {\tt sin\_t} and {\tt
  cos\_t}  in terms  of time  are plotted  in Figure~\ref{fig:cos}(b).
Figure~\ref{fig:cos}(c) shows  {\tt sin\_t} values  with corresponding
{\tt  cos\_t}  which form  almost  perfect  circle.   These plots  are
rendered    using    value    monitoring   features    described    in
Section~\ref{sec:gui}.

\renewcommand{\P}{\ensuremath{{\mathbb P}}\xspace}
\newcommand{\M}{M}

\section{Properties and Queries}
\label{sec:queries}

For  specifying   properties  of  NPTAs,  we   use  weighted  temporal
properties over  runs expressed in the  logic \WMITL \cite{mitl_2012}
(\emph{Weighted Metric Temporal Logic}),
defined by the grammar $\varphi::= ap \,|\,\neg \varphi \,|\,\varphi_1
\land  \varphi_2  \,|\,  \OO  \varphi \,|\,  \varphi_1  \U^x_{\leq  d}
\varphi_2$,  where $ap$  is an  atomic proposition,  $d$ is  a natural
number and $x$ is a clock. Here, the logical operators are interpreted
as  usual, and  $\OO$ is  a  next state  operator.  An  \WMITL-formula
$\varphi_1  \U^x_{\leq  d}  \varphi_2$   is  satisfied  by  a  run  if
$\varphi_1$ is  satisfied on the  run until $\varphi_2$  is satisfied,
and this will happen before the  value of the clock $x$ increases with
more than  $d$. For an NPTA  ${\M}$ we define $\P_\M(\psi)$  to be the
probability that a random run of $\M$ satisfies $\psi$.

The  problem  of   checking  $\P_{\M}(\psi)\geq  p$  ($p\in[0,1]$)  is
unfortunately  undecidable in  general \footnote{Exceptions  being PTA
  with 0  or 1 clocks.}.   For the sub-logic  of cost-bounded
reachability  problems  $\P_\M(\Diamond_{x\leq  C}\phi)\geq p$,  where
$\phi$  is a  state-predicate, $x$  is a  clock and  $C$ is  bound, we
approximate the  answer using simulation-based  algorithms known under
the name  of statistical model checking algorithms.   We briefly recap
statistical algorithms permitting to  answer the following three types
of questions:
\begin{enumerate}
\item  {\sl  Hypothesis Testing:}  Is  the  probability  $\P_{\M}
    (\Diamond_{x\leq  C}\phi)$ for  a given  NPTA $M$
       greater or equal to a certain threshold $p\in[0,1]$ ?
\item
{\sl Probability evaluation:} What is the probability $\P_M
  (\Diamond_{x\leq  C}\phi)$  for a given NPTA $M$?
\item 
{\sl Probability comparison:} Is the probability $\P_M
  (\Diamond_{x\leq C}\phi_2)$ greater than the probability $\P_M
  (\Diamond_{y\leq D}\phi_2]$?
\end{enumerate}

From a conceptual point of  view solving the above questions using SMC
is simple.   First, each run of  the system is encoded  as a Bernoulli
random variable  that is  true if the  run satisfies the  property and
false otherwise.  Then a statistical algorithm groups the observations
to answer the  three questions.  For the qualitative  questions (1 and
3),  we  shall  use  sequential  hypothesis  testing,  while  for  the
quantitative  question (2) we  will use  an estimation  algorithm that
resemble the  classical Monte Carlo simulation. The  two solutions are
detailed hereafter.

\paragraph{\bf  Hypothesis Testing}
This approach reduces the qualitative question to testing the hypothesis
$H: p=\P_{\boldsymbol{\M}}(\Diamond_{x\le C}\phi) \ge
\theta$ against $K: p < \theta$. To bound the probability of making
errors, we use strength parameters $\alpha$ and $\beta$ and we test
the hypothesis $H_0: p \ge p_0$ and $H_1: p \le p_1$ with
$p_0=\theta+\delta_0$ and $p_1=\theta-\delta_1$. The interval
$p_0-p_1$ defines an indifference region, and $p_0$ and $p_1$ are used
as thresholds in the algorithm. The parameter $\alpha$ is the
probability of accepting $H_0$ when $H_1$ holds (false positives) and
the parameter $\beta$ is the probability of accepting $H_1$ when $H_0$
holds (false negatives). The above test can be solved by using Wald's
{\em sequential hypothesis testing}\,\cite{Wal04}.  This test computes
a proportion $r$ among those runs that satisfy the property. With
probability 1, the value of the proportion will eventually cross
$\log(\beta/(1-\alpha)$ or $\log((1-\beta)/\alpha)$ and one of the two
hypothesis will be selected. In \uppaalsmc we use the following query:
\texttt{Pr[}$bound$\texttt{](}$\phi$\texttt{)>=}$p_0$, where $bound$
defines how to bound the runs. The three ways to bound them are 1)
implicitly by time by specifying \texttt{<=}$M$ (where M is a positive
integer), 2) explicitly by cost with $x$\texttt{<=}$M$ where $x$ is a
specific clock, or 3) by number of discrete steps with
\texttt{\#<=}$M$.  In the case of hypothesis testing $p_0$ is the
probability to test for.  The formula $\phi$ is either \texttt{<> }$q$
or \texttt{[] }$q$ where $q$ is a state predicate.

\paragraph{\bf Probability Estimation}
This algorithm~\cite{HLMP04} computes the number of runs needed in
order to produce an approximation interval
${\lbrack}p-\epsilon,p+\epsilon{\rbrack}$ for $p=Pr(\psi)$ with a
confidence $1-\alpha$. The values of $\epsilon$ and $\alpha$ are
chosen by the user and the number of runs relies on the
Chernoff-Hoeffding bound. In \uppaalsmc we use the following query:
\texttt{Pr[}$bound$\texttt{](}$\phi$\texttt{)}

\paragraph{\bf  Probability  Comparison}   This  algorithm,  which  is
detailed in  \cite{dllmpvw11}, exploits  an extended Wald  testing. In
\uppaalsmc,       we        use       the       following       query:
\texttt{Pr[}$bound_1$\texttt{](}$\phi_1$\texttt{)>=}
\texttt{Pr[}$bound_2$\texttt{](}$\phi_2$\texttt{)}.

In addition to those three classical tests, \uppaalsmc also supports
the evaluation of expected values of min or max of an expression that
evaluates to a clock or an integer value. The syntax is as follows:
\texttt{E[}$bound; N$\texttt{](min:}$expr$\texttt{)} or
\texttt{E[}$bound; N$\texttt{](max:}$expr$\texttt{)}, where $bound$ is
as explained in this section, $N$ gives the number of runs explicitly,
and $expr$ is the expression to evaluate. For this property, no
confidence is given (yet).

\paragraph{\bf  Full \WMITL} Regarding  implementation,  the reader
shall  observe  that both  of  the  above  statistical algorithms  are
trivially  implementable.  To  support  the full  logic  of \WMITL  is
slightly  more complex  as  our  simulation engine  needs  to rely  on
monitors for such logic. In \cite{mitl_2012}, we proposed an extension
of \uppaalsmc that can handle  arbitrary formulas of \WMITL.  Given a
property $\varphi$, our  implementation first constructs deterministic
under- and over-approximation monitoring  PTAs for $\varphi$.  Then it
puts these  monitors in  parallel with a  given model ${\M}$, and
applies SMC-based  algorithms to bound the  probability that $\varphi$
is satisfied on $\M$.

%\input{synsem} % moved to formalism, done
%\input{smc} % moved to queries
%\input{tool} 
%    properties/queries moved to queries
%    visualisation moved to gui
%    distributed smc moved to engine
%    done
\section{Graphical User Interface}
\label{sec:gui}
Besides short 'yes' or 'no' answers and probability estimates, \uppaalsmc verifier also provides a few statistical measures in terms of time (or cost), including frequency histogram, average time (or cost), probability density distribution, cumulative probability distribution~(the last two with confidence intervals, e.g. using the Clopper-Pearson method~\cite{Clopper34}).

\begin{figure}[htb]
\centering
\includegraphics[width=0.7\linewidth]{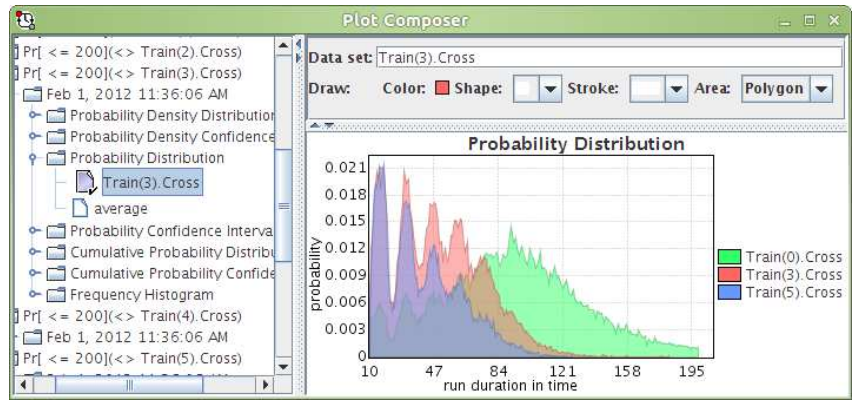}
\caption{Snapshot of the plot composer displaying three probability distributions.}
\label{fig:tgplot1}
\end{figure}

These statistical data can also be superposed onto a single plot for comparison purposes using the plot composer tool.
Figure~\ref{fig:tgplot1} shows the superposed probability distributions of trains 0, 3 and 5 crossing from our train-gate example.
On the left side of the plot composer window the user can select a particular data to be added to the plot and on the right side the user can see the superposed plot and can also change some details such as labels, shapes and colors.

\begin{figure}[htb]
\centering
\includegraphics[width=0.65\linewidth]{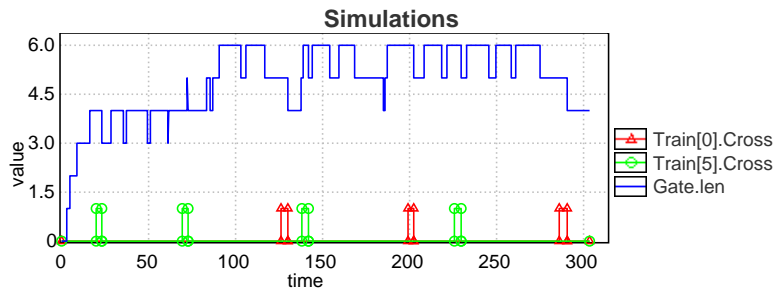}
\caption{Visualizing the gate length and when \texttt{Train(0)} and \texttt{Train(5)} cross on one random run.}
\label{fig:tg}
\end{figure}

\paragraph{Monitoring Expressions}\label{monitoring_sec}
\uppaalsmc now allows the user to visualize the values of expressions (evaluating to integers or clocks) along runs. 
This gives insight to the user on the behavior of the system so that more interesting properties can be asked to the model-checker.
To demonstrate this on our previous train-gate example, we can monitor when \texttt{Train(0)} and \texttt{Train(5)} are crossing as well as the length of the queue. 
The query is \texttt{simulate 1 [<=300]\{Train(0).Cross,Train(5).Cross,Gate.len\}}. 
This gives us the plot of Figure~\ref{fig:tg}. 
Interestingly \texttt{Train(5)} crosses more often (since it has a higher arrival rate). 
Secondly, it seems unlikely that the gate length drops below 3 after some time (say 20), which is not an obvious property from the model. 
We can confirm this by asking \texttt{Pr[<=300](<> Gate.len < 3 and t > 20)} and adding a clock \texttt{t}. The probability is in $[0.102,0.123]$.

\begin{figure}[htb]
\centering
\begin{tabular}{cccc}
  \begin{minipage}[b]{0.14\linewidth}
    \centering
    \includegraphics[width=\linewidth]{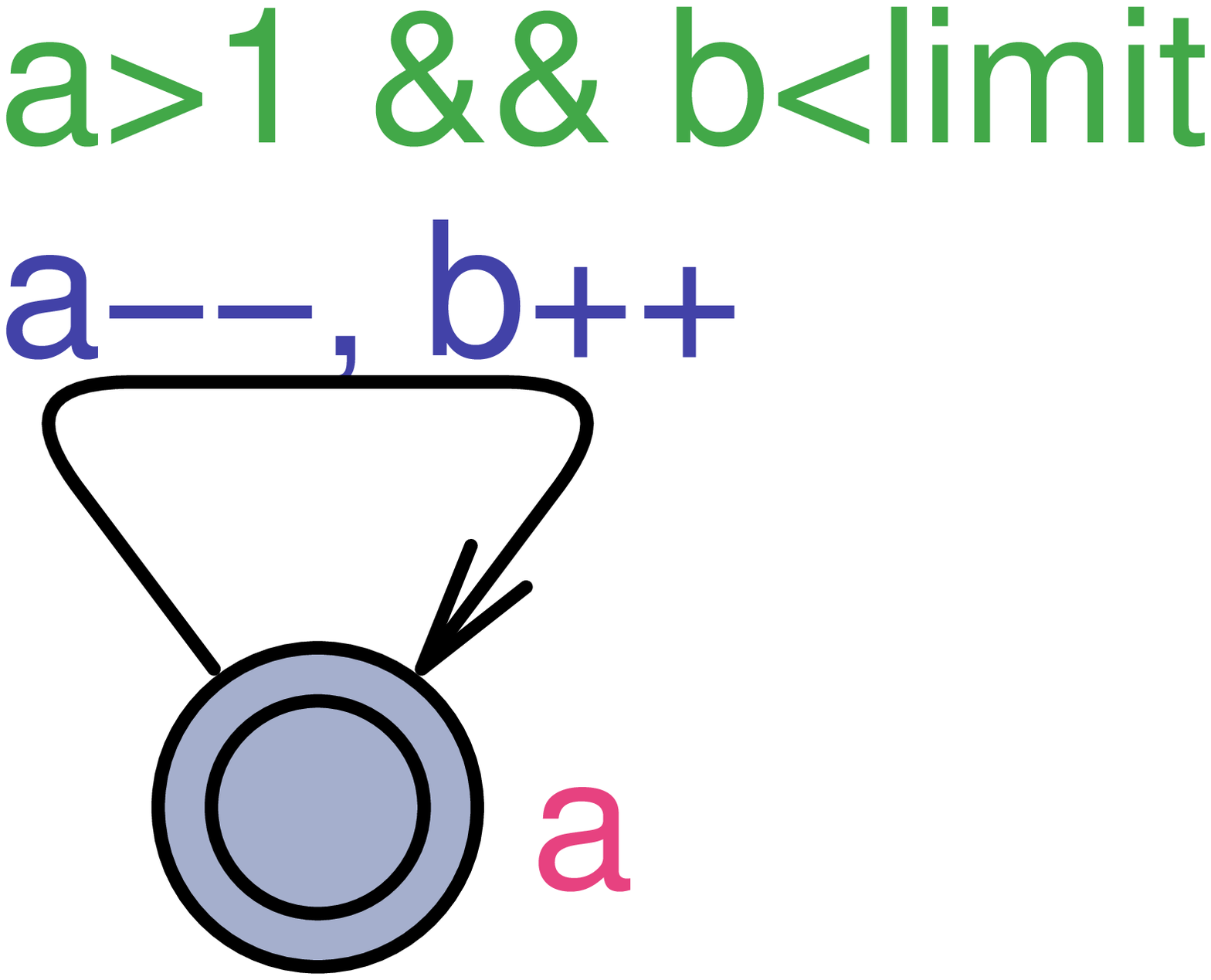}\\
    (a)\medskip\\
    \includegraphics[width=\linewidth]{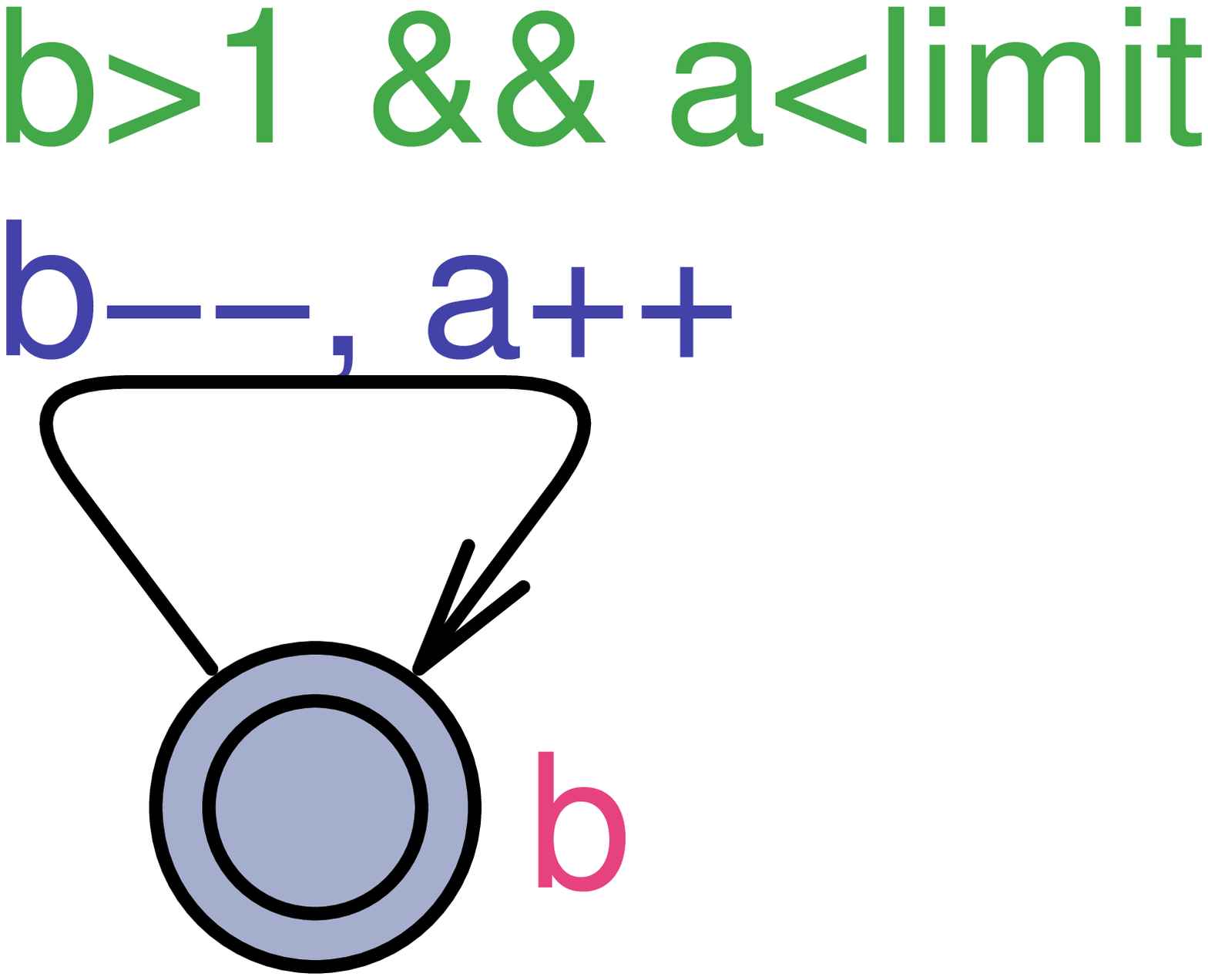}
  \end{minipage} &
  \includegraphics[height=0.18\textheight]{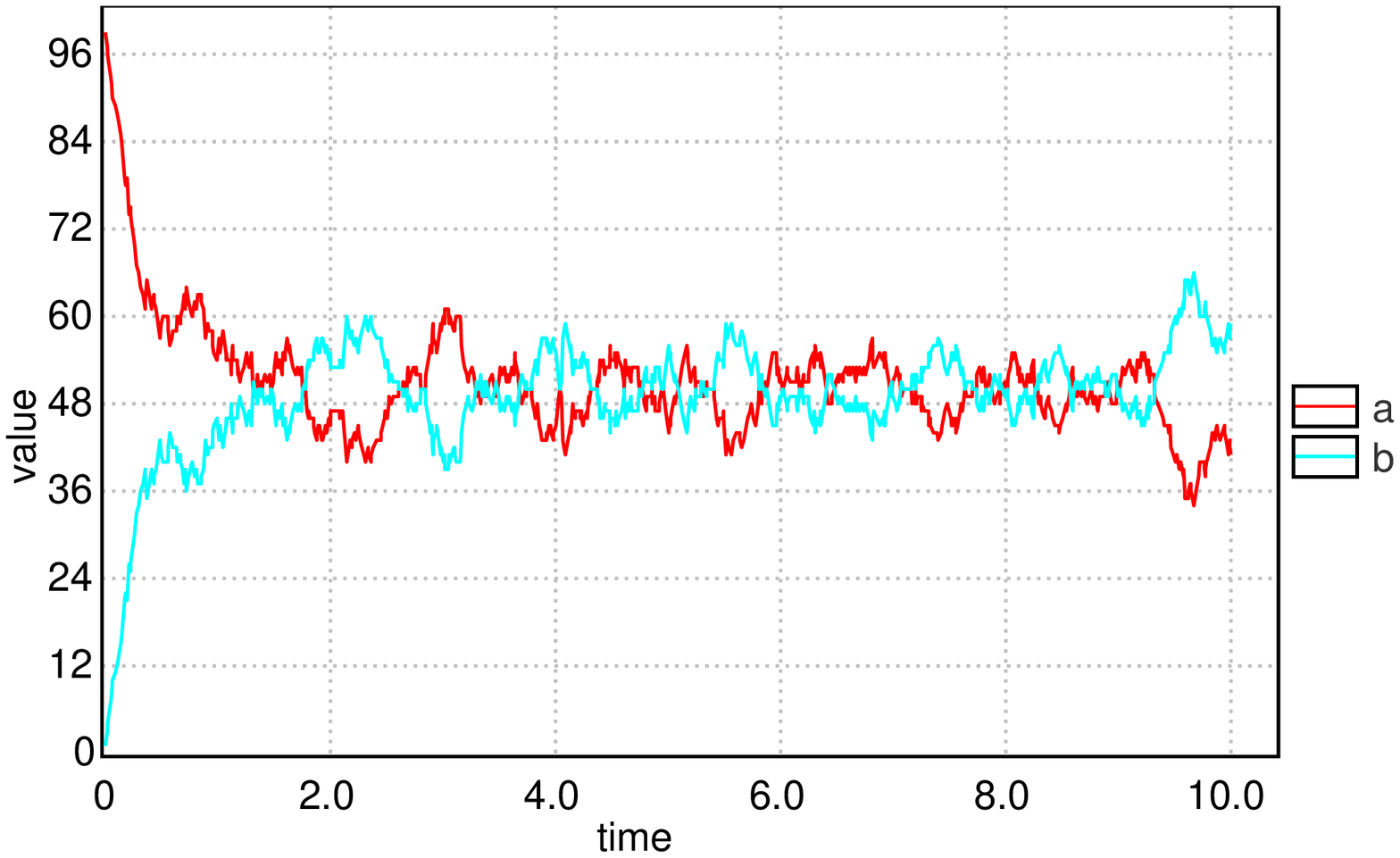} &
  \includegraphics[height=0.18\textheight]{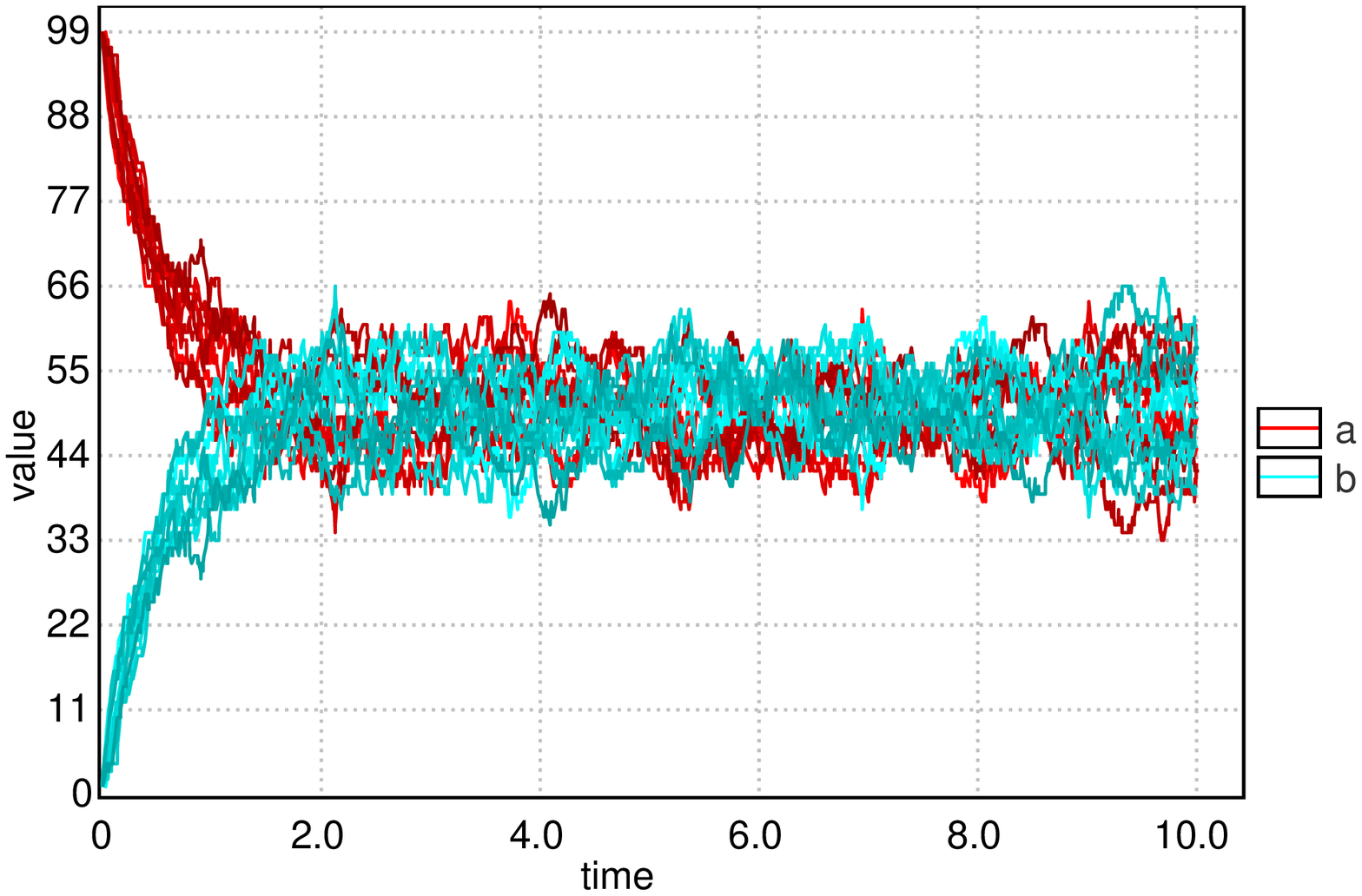} \\
   (b) & (c) & (d)
\end{tabular}
\caption{Evolution of the concentrations of two reactants \texttt{a} and \texttt{b}.}
\label{fig:bio}
\vspace{-5pt}
\end{figure}

As a second example to illustrate this feature, we consider the modeling of chemical reactions. 
Figure~\ref{fig:bio}(a) and \ref{fig:bio}(b) show two symmetric timed automata that model the concentrations of reactants \texttt{a} and \texttt{b} (here as integers). 
The exponential rate for taking the transition is given by the concentration of \texttt{a} and \texttt{b}. 
Figure~\ref{fig:bio}(c) shows the evolution of the system when it is started with \texttt{a=99} and \texttt{b=1}: \texttt{a} is consumed to produce \texttt{b} and vice-versa, and the concentrations oscillate.

The   simulations  are   obtained  by   querying   \texttt{simulate  1
  [<=10]\{a,b\}}. Figure~\ref{fig:bio}(c) is  showing one evolution of
\texttt{a} and \texttt{b} over time.  The tool can also plot clouds of
trajectories, which is useful to identify patterns in the behavior, as
shown in figure~\ref{fig:bio}(d).

It is important to notice that generating such curves is not as
trivial as it seems. In fact, on such models, if the exponential rates
are higher, then the time steps are much smaller, which generates a
lot of points, up to consuming several GB of memory. Drawing such
plots is not practical. The tool would not work due to out-of-memory
problems or in the best case will take around 30s to transfer the data
and several seconds for every redraw. To solve this the engine applies
an on-the-fly filtering of the points based on the principle that if
two points are too close to each other to be distinguished on the
screen, then they are considered to be the same. A resolution
parameter is used to define the maximal resolution of the plot and
eliminates the memory and speed problems completely (down to almost
not measurable).

This plot in Figure~\ref{fig:cos}(b) is obtained by asking \texttt{simulate 1 [<=12]\{sin\_t,cos\_t\}} to the model-checker. 
Interestingly, \uppaalsmc can generate a run bounded by any clock so we can also plot \texttt{simulate 1 [cos\_t<=1]\{sin\_t\}} and obtain a circle as shown in Figure~\ref{fig:cos}(c). 

\section{Engine}
\label{sec:engine}

The actual techniques to achieve the current performance of the tool
were never exposed before. In this section, we present a few key
optimizations to implement the algorithms presented and new features
that were not available in earlier versions of \uppaalsmc.

\vspace{-5pt}
\paragraph{Distributed SMC}\label{distr_descr}
The problem in distributing the implementation of the sequential SMC
algorithm is that a \emph{bias} may be introduced. The reason is that
sequential testing relies on collecting outcomes of the generated runs
on-the-fly. If some computation cores generate some accepting runs faster,
which is possible if rejecting runs happen to be longer or simply more
expensive to compute, then the result will be biased. The solution of
this problem is to force all the cores to generate the same amount
of simulations.
The paper \cite{You05a} proposes a method to ensure this by splitting
the simulations into batches of the same size, and this method has
been generalized and implemented in \uppaalsmc~\cite{bdlml11}. The
distributed implementation gives a linear speed-up in the number of
cores used.

\vspace{-5pt}
\paragraph{Detection of States}
When choosing the delays, the engine does not know if it will
\emph{skip} the state that should be observed by the query or
not. This problem is present when picking delays to take transitions
as well. For example, the query could be \texttt{<> A.critical and x >= 2 and x <= 3} where \texttt{x} is a clock. The engine should not delay 4 time units from a state where \texttt{x=0} because the first possible transition is enabled at this point. Special care is taken to make sure that the formula is part of the next \emph{interesting} points that are computed when choosing the delays. Now comes the question of how to detect those interesting points in both the formula and the guards.

The technique we use follows the decorator pattern where we evaluate guards (for detecting which transitions will be enabled in the future) and formulas in the query to keep track of the lower bounds. We wrap a state inside a decorator state that keeps track of the constraints on-the-fly, only remembering the bounds that we need. The point of the technique here is to avoid \emph{symbolic} states that would require zones typically implemented with different bound matrices.

\vspace{-5pt}
\paragraph{Early Termination}
The engine checks for query on-the-fly on every generated run. If a
query is satisfied then the computation of the run is stopped before
it reaches the specified bound. In addition, in order to give the user
a way to stop runs earlier, the engine supports an \emph{until}
property: \texttt{p U q} can be queried instead of \texttt{<> q} and
cut the runs as soon as \texttt{p} stops to hold.

\paragraph{Dependencies and Reuse of Choice}
When a  process takes  an action, it  may not affect  other processes,
which means that from a  stochastic point-of-view, picking a new delay
from  scratch  or  \emph{reusing}  the   old  (random) choice is
equivalent.  The  engine exploits  this independence: it  remembers the
previous  delays chosen  by the  processes and  invalidates  them when
dependent transitions  are taken. A process has  its delay invalidated
if there  is a dependency  with another transition being  taken, which
happens in  case of  synchronization or a  dependency through  a clock
rate, invariant, guard,  or update.  A static analysis  is made at the
granularity  of  \emph{how  transitions affect  processes}\footnote{We
  judge that keeping  track of the dependencies down  to the locations
  may have a too large overhead.}.

The result is that whenever a process \emph{needs} to pick a delay, it does so. Whenever a process takes a transition, the processes that may be affected by it must pick a new delay at the next step. Otherwise, processes \emph{reuse} their choices from the previous step in the simulation\footnote{If time elapses then of course the delays chosen are updated.}.

Checking the query \texttt{Pr[<=300](<> Train(0).Cross and (forall
  (i:id\_t) i!=0 imply}
\begin{wraptable}{r}{0.6\linewidth}
\centering
\vspace{-5pt}
\begin{tabular}{|l|r|r|r|r|}\hline
Trains   & 5             & 10            & 20        & 40 \\ \hline
Proba.   & 0.985-0.995 & 0.286-0.297 & 0-0.008 & 0-0.005 \\
Time$^-$ & 3.9s          & 17.3s         & 41.1s     &  98.1s \\
Time$^+$ & 3.5s          & 14.8s         & 33.2s     &  74.8s \\
Gain     & 10.2\%        & 14.4\%        & 19.2\%    &  23.8\% \\ \hline
\end{tabular}
\caption{Probability and time results without (-) and with (+) reuse.}
\vspace{-5pt}
\label{tab:tg}
\end{wraptable}
\noindent
\texttt{Train(i).Stop))}
 to evaluate the probability of
\texttt{Train(0)} crossing while all the others are stopped gives the
results in Table~\ref{tab:tg} for different numbers of trains. The
results are obtained with the parameter $\epsilon=0.005$ and the
probability results agree with or without reuse within $\epsilon$. The
experiments are made on a core i7 at 2.66GHz. This optimization is
designed to improve on systems with large number of components, which
is shown by the increasing improvement relative to verifications without reuse.

\section{Case-Studies}
\label{sec:cases}

In this section we evaluate the applicability of the developed techniques on practical case studies.

\paragraph{Robot Control}\label{robot}

In  paper  \cite{mitl_2012} we  considered a  case --  explored in
\cite{DBLP:conf/tacas/BarbotCHKM11}  --   of  a  robot   moving  on  a
two-dimensional grid.  Each field of  the grid is either {\tt normal},
on {\tt  fire}, cold  as {\tt  ice} or it  is a  wall which  cannot be
passed.   Also,  there is  a  {\tt goal}  field  that  the robot  must
reach. The  robot is moving  in a random  fashion i.e.  it stays  in a
field for some  time, and then moves to a  neighboring field at random
(if it is not a wall).

\begin{wrapfigure}[8]{r}{0.35\textwidth}
\vspace{-10pt}
      \includegraphics[trim=0mm 0cm 0mm 1.3cm,clip,height=0.15\textheight]{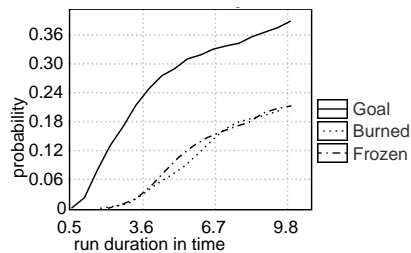}
\vspace{-12pt}  
\caption{Cumulative Probability}\label{fig:robotDistribution}
\end{wrapfigure}  

We are interested  in the probability that the  robot reaches its goal
location without staying on consecutive  fire fields for more than one
time unit and on consecutive ice fields for more than two time units.
This property is captured by the \WMITL\ formula
$\varphi   \equiv  (\varphi_1   \land  \varphi_2)   \U^\tau_{\leq  10}
\,\text{\tt goal}$, where $\tau$ is a special clock that grows with rate $1$ and is never reset, and:
\begin{small}
\begin{align*}
& \varphi_1 \equiv \text{\tt ice} \implies \Diamond^\tau_{\leq 2} (\text{\tt fire} \lor \text{\tt normal} \lor \text{\tt goal}) \\
& \varphi_2 \equiv \text{\tt fire} \implies \Diamond^\tau_{\leq 1} (\text{\tt ice} \lor \text{\tt normal}\lor \text{\tt goal}) 
\end{align*} 
\end{small} 
\,\,\, We applied \uppaalsmc to compute the probability of the robot reaching the goal $\varphi$, staying too long in the fire or too long on the ice. Figure \ref{fig:robotDistribution} shows the cumulative distribution for these probabilities.

%%% Local Variables: 
%%% mode: latex
%%% TeX-master: "main"
%%% End: 

\paragraph{Firewire.} IEEE 1394 High Performance Serial Bus or Firewire for short is used to transport multimedia signals among a network of consumer devices.
The protocol has been extensively studied (see~\cite{firewireCmp03} for comparison) and in particular~\cite{firewirePrism} uses probabilistic timed automata in {\sc Prism}~\cite{KNP04}.
In paper~\cite{dllmw11} we adopt the model from~\cite{firewirePrism} and demonstrate how \uppaalsmc can be used to evaluate fairness of a node becoming a root (leader) with respect to the mode of operation.
\begin{wrapfigure}{r}{5.6cm}
  \vspace{-10pt}
  \includegraphics[width=5.5cm]{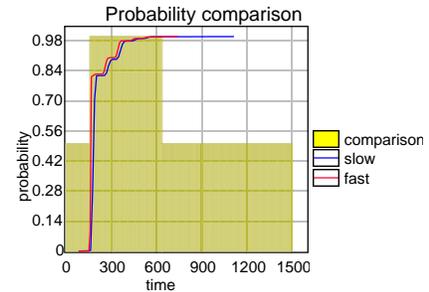}
  \vspace{-10pt}
  \caption{Probability Comparison}\label{fig:firewire}
\end{wrapfigure}
\uppaalsmc provides two methods for comparing probabilities: estimating the probabilities and then comparing them, or using indirect probability comparison from~\cite{Wal04}, which is more efficient.
Figure~\ref{fig:firewire} contains a resulting plot of estimated probabilities (red and blue lines) and a comparison (yellow area).
The red and blue probability estimates appear very close to each other in entire range, while the yellow area shows that at the beginning the probabilities are indistinguishable (yellow area is at 0.5 level), then the \emph{fast} node has higher probability to become a \emph{root} (at 1.0 level), and later the probabilities become too close to be distinguishable again (at 0.5 level).

\begin{wrapfigure}{r}{0.35\textwidth}
  \vspace{-10pt}
  \includegraphics[width=\linewidth]{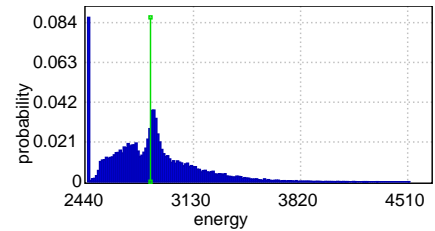}
  \vspace{-25pt}
  \caption{Energy consumption.}
  \label{fig:bluetooth}
  \vspace{-20pt}
\end{wrapfigure} %
\paragraph{Bluetooth~\cite{bluetooth}}
is a wireless telecommunication protocol using frequency-hopping to cope with 
interference between the devices in the wireless network.
In paper~\cite{dllmw11} we adopted the model from~\cite{DKNP06}, annotated the model to record the power utilization and evaluated the probability distributions of likely response times and energy consumption.
Figure~\ref{fig:bluetooth} shows that after 70s the cost of a device operation is at least $2440$ energy units and the mean is about $2853$ energy units.

\paragraph{Lightweight Medium Access Protocol (LMAC)~\cite{LMAC}} is a communication scheduling protocol based on time slot distribution for nodes sharing the same medium. 
The protocol is designed having wireless sensor networks in mind: it is simple enough to fit on a modest hardware and at the same time robust against topology reconfiguration, minimizing collisions and power consumption.
Paper~\cite{FHM2007} studies LMAC protocol using classical {\sc Uppaal} verification techniques by systematically exploring networks of up to five nodes but the state space explosion prevents formal verification of larger networks.
In paper~\cite{dllmpvw11} we adopt the model by removing verification optimizations and parameterizing with probabilistic weights, and show how collisions can be analyzed and power consumption estimated using statistical model checking techniques.
The study showed that there are still perpetual collisions in a ring topology
\begin{wrapfigure}{r}{0.48\textwidth}
  \vspace{-10pt}
  \includegraphics[width=0.48\textwidth]{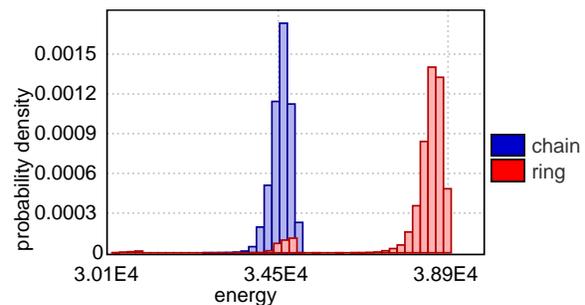}
  \vspace{-20pt}  
  \caption{Likely energy consumption.}
  \label{fig:lmac-power}
  \vspace{-10pt}  
\end{wrapfigure} %
but the probability that the network will not recover is very low (0.35\%).
The likely energy consumption of different network topologies is compared in {\sc Uppaal} plot (Figure~\ref{fig:lmac-power}), which shows that on average the likely energy consumption after 1000 time units in a ring is higher than in a chain by $10\%$, possibly due to more collisions in a ring.
In~\cite{bdlml11} distributed techniques are applied in exploring over 10000 larger networks of up to 10 nodes, the worst (star-like) and the best (chain-like) topologies in terms of collisions are identified and evaluated.

\paragraph{Computing Nash Equilibrium in Wireless Ad~Hoc Networks}

One of the important aspects in designing wireless ad-hoc networks is
to make sure that a network is robust to the selfish behavior of its
participants, i.e. that its configuration satisfies Nash equilibrium~(NE).

%We assume that
%each node can work according to one out of finitely many
%configurations that we call strategies.  We also assume that each network node has a goal and
%a node's utility function is equal to the probability that this goal
%will be reached on a random system run. 

\begin{wrapfigure}{r}{0.5\textwidth}
  \vspace{-35pt}
  \includegraphics[width=0.48\textwidth]{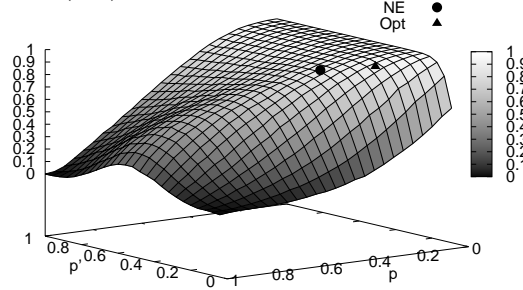}
  \vspace{-35pt}  
  \caption{Nash Equilibrium for Aloha CSMA/CD}\label{nash_fig}
  \vspace{-10pt}  
\end{wrapfigure}
In paper \cite{nash_2012} we proposed an SMC-based algorithm for computing NE for the case when network nodes are modeled by SPTA and an utility function of a single node is equal to a probability that the node will reach its goal.
Our algorithm consists of two phases.
First, we use \uppaalsmc to find a strategy that most likely (heuristic) satisfies NE. 
In the second phase we apply statistics to test the
hypothesis that this strategy actually satisfies NE. 

We applied this algorithm to compute NE for  Aloha CSMA/CD and IEEE 802.15.4 CSMA/CA protocols.
Figure~\ref{nash_fig} depicts the utility function plot for the Aloha CSMA/CD protocol with two nodes. 
Here the $p$ and $p'$ axis correspond to the strategies of the honest and cheater nodes (a strategy defines how persistent these nodes are in sending their data).
We see, that NE strategy is slightly less efficient than the symmetric optimal strategy (Opt), but it still results in a high value of the utility function.

%%% Local Variables: 
%%% mode: plain-tex
%%% TeX-master: "main"
%%% End: 

% Danny
\paragraph{Duration Probabilistic Automata}\begin{wrapfigure}{r}{0.53\textwidth}
  \vspace*{-3mm}
  \includegraphics[width=\linewidth]{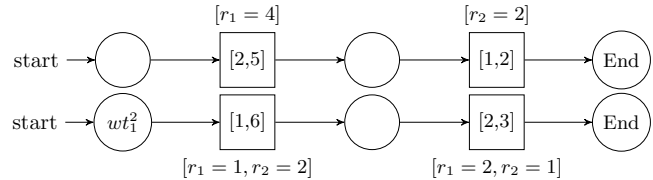}
  \vspace*{-8mm}
  \caption{Rectangles are busy states and circles are for waiting when resources are not available. There are $r_1=5$ and $r_2=3$ resources available.}
  \label{fig:DPAexample}
  \label{fig:dpa}
  \vspace*{-20pt}
\end{wrapfigure}
In \cite{uppaal_formats} we compared \uppaalsmc to Prism \cite{KNP04} in the context of Duration Probabilistic Automata (DPA) \cite{MLK10}.
A Duration Probabilistic Automaton (DPA) is a composition of Simple Duration Probabilistic Automata (SDPA). An SDPA is a linear sequence of tasks that must be performed in a sequential order.
\newcommand{\upp}{\ensuremath{Up}\xspace}
\begin{wraptable}{r}{0.6\textwidth} %[htbp]
  % \centering
  \vspace{-5pt}
  %\vspace{-25pt}
  %The results are from the formats 2011 paper 
 {\footnotesize
  \begin{tabular}{@{\extracolsep{-1pt}}*{3}{r}|*{4}{r}|*{4}{r}}
  \toprule
    \multicolumn{3}{c|}{\bf Param.} &   \multicolumn{4}{c|}{\bf Estim.} & \multicolumn{4}{c}{\bf Hyp. Testing} \\ 
    $n$ & $k$ & $m$  &Prism & $\upp_p$ & $\upp_d$ & $\upp_c$ & Prism & $\upp_p$ & $\upp_d$ & $\upp_c$ \\ 
    \hline
    
4 & 4 & 3& 2.7 & 0.3 & 0.2 & 0.2  & 2.0  & 0.1 & 0.1 & 0.1 \\
6 & 6 & 3& 7.7 & 0.6 & 0.5 & 0.4  & 3.9  & 0.2 & 0.2 & 0.3 \\
8 & 8 & 3& 26.5 & 1.2 & 0.9 & 0.7 & 16.4 & 0.5 & 0.4 & 0.3\\
    \hline

    20 & 40&20&   \multicolumn{4}{c|}{$>$300} & $>$300 & 35.5 & 26.2 & 20.7 \\  
    30 & 40& 20&   \multicolumn{4}{c|}{$>$300}& $>$300 & 61.2 & 41.8 & 33.2\\
    40 & 40& 20&  \multicolumn{4}{c|}{$>$300} & $>$300 & 92.2 & 56.9 & 59.5 \\ 
    40 & 20& 20&  \multicolumn{4}{c|}{$>$300} & $>$300 & 41.1 & 31.2 & 26.5 \\ 
    40 & 30 &20&  \multicolumn{4}{c|}{$>$300}  & $>$300& 68.8 & 46.7 & 46.1\\ 
    \hline

    40 & 55 & 40&\multicolumn{4}{c|}{$>$300}  & \multicolumn{3}{c}{$>$300} & 219.5 \\ 
    \bottomrule
  \end{tabular}}
\caption{Performance of SMC (sec). The $n$ column is the number of SDPAs, the $k$ column is the number of tasks per SDPA and the $m$ column is the number of resource types in the model. $\upp_p$ is the \uppaal model that matches Prism, $\upp_d$ the discrete encoding and $\upp_c$ the continuous time encoding.}\label{tab:uppprismcomp}
\vspace{-10pt}
\end{wraptable}

Each task is associated with a duration interval which gives the possible durations of the task. The actual duration of the tasks is given by a  uniform choice from this interval. To model races between the SDPAs we introduce resources to the model such that an SDPA might have to wait for resources before processing a task. When two SDPAs are in waiting position for the same resource, a scheduler decides which SDPA is given the resource in a deterministic manner.

 The comparison with Prism was made by randomly generating models with a specific number of SDPAs and a specific number of tasks per SDPA and translate these into Prism and \uppaal models. The Prism model uses a discrete time semantics whereas three models were made for \uppaal - one with continuous time semantics, one that matches the Prism model as close as possible and one with discrete semantics that makes full use of our formalism.  \\
The  queries to the models were \textit{What is the probability of all SDPAs ending within $t$ time units} (Estimation)and \textit{Is the probability that all SDPAs end within $t$ time units greater than $40\%$} (Hypothesis testing). The value of $t$ is different for each model as it was computed by simulating the system 369 times and represent the value for which at least $60\%$ of the runs finished all their tasks.

The result of the experiments are shown in Table \ref{tab:uppprismcomp} and indicates that \uppaal is notably faster than Prism, even with a encoding that closely matches that of Prism.

%%% Local Variables: 
%%% mode: latex
%%% TeX-master: "main"
%%% End: 

\begin{wrapfigure}{r}{0.42\textwidth}
  \vspace{-5pt}
  \includegraphics[width=0.42\textwidth]{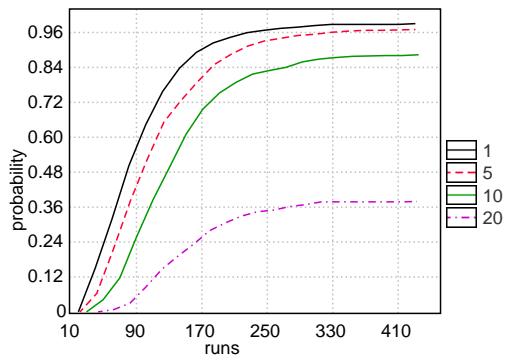}
  \vspace*{-15pt}
  \caption{Probability distributions obtained with 1, 5, 10, and 20 cores.}
  \label{dist_fig}
  \vspace{-8pt}
\end{wrapfigure}
\paragraph{Checking of Distributed Statistical Model Checking}
As we wrote in Section~\ref{distr_descr}, a naive (and incorrect) distributed implementation of the sequential SMC algorithms might introduce a bias towards the results that are generated by shorter simulations. 

The interesting question is how much this bias affects the SMC results.
In the paper~\cite{nfm_2012} we answered this question by modeling the naive distributed SMC algorithm in \uppaalsmc\ itself.
The comparison was made on the basis of the SPTA model that ends up in the \verb|OK| location after $100$ time units with probability $0.58$, otherwise it ends up in the \verb|NOK| location after $1$ time unit (thus producing \verb|NOK| requires $100$ times less time than producing \verb|OK|).

We used \uppaalsmc\ to compute the probability that the naive distributed SMC algorithm will accept the hypothesis {\tt Pr[<=100]($\Diamond$  OK)$\geq$ 0.5}. The results for the different numbers of computational cores are given in the plot at Figure~\ref{dist_fig}.
The $x$ axis denotes the total number of runs of the SPTA model on all the cores, and the $y$ axis depicts the probability that an SMC algorithm accepts the hypothesis not later than after this number of runs.
It can be observed that the probability of accepting the hypothesis tends~(incorrectly) to $0$ as the number of computational cores increases.
%\uppaalsmc doesn't have this problem since it forces all the cores to perform the same number of simulations. 

%The {\sc Uppaal} implementation of the distributed SMC algorithm doesn't have this problem, as it has been described in~\cite{You05a}.

%\input{future}
\section{Conclusions}
\label{sec:conclusion}

This paper gives an overview of the features of \uppaalsmc, our new efficient
extension of \uppaal for Statistical Model Checking. Contrary to other
existing SMC-based tool-sets, \uppaalsmc allows to handle
systems with real-time features. The tool has been applied to a series
of case studies that are beyond the scope of classical model
checkers. As has been outlined in this paper, \uppaalsmc has a large
potential for future work and applications.

Among others, the following extensions of \uppaalsmc are contemplated.

\paragraph{Floating Point}
So far the support of floating point is done via misusing and extending clock operations. A better and more general support is needed since the tool has now departed from traditional timed automata and model-checking.

Since the tool now supports floating point arithmetic and we can integrate complex functions, it is a natural extension to add differential equations as well to support hybrid systems in a more general way. To fit with the stochastic semantics (in particular how to pick delays), only simple equations whose analytical solutions are known are planned.

\paragraph{New Applications}
With the extended expressivity of our hybrid modeling language, our tool can be applied to different domains, in particular for biological systems. \uppaalsmc now offers powerful visualization capabilities needed by biologists and a logic to do statistical model-checking.

Another application is to analyze performance of controllers generated by \tiga~\cite{BCDFLL07}, in particular their stability or energy consumption. SMC can also be used in the domain of refinement checking, which is in the end just another type of game.

\paragraph{Rare Events}
Statistical model checking avoids the exponential growth of states
associated with probabilistic model checking by estimating properties
from multiple executions of a system and by giving results within
confidence bounds. Rare properties are often very important but pose a
particular challenge for simulation-based approaches, hence a key
objective under these circumstances is to reduce the number and length
of simulations necessary to produce a given level of
confidence. Importance sampling is a well-established technique that
achieves this, however to maintain the advantages of statistical model
checking it is necessary to find good importance sampling
distributions without considering the entire state space. Such problem
has been recently investigated for the case of discrete stochastic
systems. As an example, in \cite{JLS12} we presented a simple
algorithm that uses the notion of cross-entropy to find the optimal
parameters for an importance sampling distribution. Our Objective is
to extend our results to PTAs by exploiting pure timed model checking
to improve the search for efficient distribution.

%\nocite{*}
\bibliographystyle{eptcs}
\bibliography{references,thesis,paper,main}
\end{document}